\documentclass[aps,amssymb,amsmath,preprintnumbers,floatfix,twocolumn,nofootinbib,superscriptaddress]{revtex4}
\pdfoutput=1 
\usepackage{dcolumn}
\newcommand{\of}[1]{\left(#1\right)}

\newcommand{\lp}{\left(}
\newcommand{\rp}{\right)}
\newcommand{\lb}{\left[}
\newcommand{\rb}{\right]}

\newcommand{\abs}[1]{\left|#1\right|}

\newcommand{\bigS}{\mathbb{S}}

\def\m{\mu}

\newcommand{\be}{\begin{equation}}
\newcommand{\ee}{\end{equation}}
\newcommand{\br}{\begin{eqnarray}}
\newcommand{\bea}{\begin{eqnarray}}
\newcommand{\eea}{\end{eqnarray}}
\newcommand{\er}{\end{eqnarray}}
\newcommand{\ba}{\begin{array}}
\newcommand{\ea}{\end{array}}
\newcommand{\bi}{\begin{itemize}}
\newcommand{\ei}{\end{itemize}}
\newcommand{\bn}{\begin{enumerate}}
\newcommand{\en}{\end{enumerate}}
\newcommand{\bc}{\begin{center}}
\newcommand{\ec}{\end{center}}

\newcommand{\beq}{\begin{equation}}
\newcommand{\eeq}{\end{equation}}

\newcommand{\gsim}{\lower.7ex\hbox{$\;\stackrel{\textstyle>}{\sim}\;$}}
\newcommand{\lsim}{\lower.7ex\hbox{$\;\stackrel{\textstyle<}{\sim}\;$}}
\usepackage{amsfonts}
\usepackage{amssymb}
\usepackage{amsmath}
\usepackage{amssymb}
\usepackage{wasysym}
\usepackage{bm}
\usepackage{latexsym}
\usepackage{textcomp}
\usepackage{verbatim}
\usepackage{slashed}
\usepackage{array}
\usepackage{graphicx}
\usepackage{multirow}
\usepackage{graphicx}

\usepackage{color}
\usepackage{enumitem}

\begin{document}

\title{Impact of a complex singlet: electroweak baryogenesis and dark matter}

\author{Minyuan Jiang}
\affiliation{Department of Physics, Nanjing University, 22 Hankou Road, Nanjing 210098, P. R. China}
\affiliation{State Key Laboratory of Theoretical Physics and Kavli Institute for Theoretical Physics China, Institute of Theoretical Physics, Chinese Academy of Sciences, Beijing 100190, P. R. China}
\author{Ligong Bian}
\affiliation{State Key Laboratory of Theoretical Physics and Kavli Institute for Theoretical Physics China, Institute of Theoretical Physics, Chinese Academy of Sciences, Beijing 100190, P. R. China}
\affiliation{Department of Physics, Chongqing University, Chongqing 401331, P.R. China}
\author{Weicong Huang}
\affiliation{State Key Laboratory of Theoretical Physics and Kavli Institute for Theoretical Physics China, Institute of Theoretical Physics, Chinese Academy of Sciences, Beijing 100190, P. R. China}
\author{Jing Shu}
\affiliation{State Key Laboratory of Theoretical Physics and Kavli Institute for Theoretical Physics China, Institute of Theoretical Physics, Chinese Academy of Sciences, Beijing 100190, P. R. China}


\date{\today}

\date{\today}

\begin{abstract}

With the assistance of a complex singlet, and an effective operator involving CP violations,
the dark matter relic abundance and baryon asymmetry of the universe
have been addressed simultaneously. We studied the electroweak baryogenesis mechanism systematically.
The electroweak phase transition analysis indicates that the strong first order phase transition
takes place by one-step or two-step type due to the dynamics of the energy gap between
the electroweak vacuum and the vacuum of the complex singlet. The relation between the magnitude of baryon asymmetry of the universe and the phase transition type and strength has been explored in the framework of electroweak baryogenesis.

\end{abstract}

\maketitle

\section{Introduction}
\label{sec:intro}

The Standard Model (SM) has passed most experimental tests during the last 40 years.
And the discovery of a 125 GeV SM-like Higgs particle by the ATLAS~\cite{Aad:2012tfa} and CMS~\cite{Chatrchyan:2012ufa} collaborations at the LHC seems to provide the last missing piece of the SM.
However, it has long been known that the SM has two obvious shortcomings, i.e., the explanations of the observed dark matter abundance and
matter-anti-matter asymmetry of the universe.

Firstly,
the observed dark matter abundance is one issue that couldn't be addressed in the SM.
The existence of dark matter (DM) has already been established by
the observations of galaxy rotation curves and analysis of cosmic microwave background (CMB) etc.
And the PLANCK~\cite{Ade:2013zuv} and WMAP~\cite{Jarosik:2010iu} predict that DM constitutes about 26.5\% of our Universe.
Secondly, the matter-anti-matter asymmetry of the universe,
i.e., baryon asymmetry in the Universe (BAU) as a baryon to entropy ratio~\cite{Ade:2013zuv,Beringer:1900zz}
 \begin{eqnarray}
\frac{n_b}{s}\approx(0.7-0.9)\times10^{-10}\; ,
\end{eqnarray}
 couldn't be addressed and predicted in the framework of the SM.
 The dynamical generation of BAU requires three necessary ingredients~\cite{Sakharov:1967dj}:
   (1) violation of baryon number; (2) violation of both C- and CP-symmetry; and (3) departure from equilibrium dynamics or CPT violation.
Though the SM contains all these requirements, and the electroweak phase transition (EWPT) provides a natural mechanism for baryogenesis~\cite{Sakharov:1967dj}, the SM is unable to solve the problem for the reasons being listed below:
Although the baryon number violation is provided in the SM by weak sphalerons~\cite{Klinkhamer:1984di,Gavela:1994dt,Huet:1994jb},
the departure from thermal equilibrium, provided by a strong first order phase transition, proceeding by bubble nucleation which makes
electroweak symmetry breaking (EWSB), does not occur in the SM~\cite{Aoki:1999fi}.
In the SM, no first order phase transition occurs for Higgs mass larger than about 80 GeV~\cite{Kajantie:1996mn,Kajantie:1996qd,Csikor:1998eu},
which is far below the experimental bound of $m_h$ $>$ 114 GeV from LEP ~\cite{LEP:2003aa} and $m_h$$\approx$125 GeV from the LHC~\cite{Aad:2012tfa,Chatrchyan:2012ufa}.
 In addition, the CP violation in the CKM matrix is too small to produce a sufficiently large baryon number.
New CP violations beyond the SM are required~\cite{Konstandin:2003dx}.

This work is aimed at accommodating the
observed DM relic density and the BAU simultaneously.
The observed DM relic density reported by CMB anisotropy probes suggests weakly interacting massive
particle (WIMP)~\cite{Griest:2000kj,Bertone:2004pz} and a feasible candidate for DM requires the
extension of the SM.
Based on $CxSM$~\cite{Barger:2008jx}, we consider one simple extension of the SM
with a complex singlet ($\bigS$) being supplemented. The singlet transforms trivially under the SM gauge groups and the imaginary part takes the responsibility of being DM.
One very attractive mechanism, i.e., electroweak baryogenesis (EWBG),
wherein baryon number generation is driven by the CP asymmetry
at the time of the EWPT\footnote{Which should be the strong first-order EWPT to protect the
generated baryon asymmetry from washout and suppress the sphaleron after the process is ended.}, is adopt.
We take advantage of the complex singlet $\bigS$ to obtain the CP-violation required by EWBG through introducing a dimension-6
operator.
Ref.~\cite{Espinosa:2011eu} considered the
analogous dimension-5 operator involving $S/\Lambda$, and Ref.~\cite{Cline:2012hg}  use $S^2/\Lambda^2$ because of the $Z_2$ symmetry
$S\to -S$ needed to prevent decay of $S$, as befits a dark matter candidate.  In our case, we consider the complex scalar $\bigS$, thus we have
$\bigS \bigS^\dagger/\Lambda^2$, which ensures that the dark matter candidate preserves $Z_2$ symmetry and the CP violation source survives after symmetry breaking.
The relevant Lagrangian takes the form of
\begin{equation}
	y_t \bar Q_L H \left(1+ {(a \;+ i b)\over\Lambda^2}\bigS \bigS^\dagger\right)t_R +{\rm h.c.}
\label{dim6op}
\end{equation}
where $a,~b$ are real parameters and $\Lambda$ is a new
physics scale. During the EWPT, the top quark mass  gets a
spatially-varying complex phase along the bubble wall profile,
which provides the source of CP violation needed to generate the
baryon asymmetry.
Precision tests, especially the electric dipole moment (EDM) searches, can probe directly CP violation relevant to EWBG.
Using the polar molecule thorium monoxide (ThO), the ACME collaboration reported an upper limit on the electron EDM (eEDM) recently~\cite{Baron:2013eja}, at 90$\%$ confidence level, an order of magnitude stronger than the previous best limit,
\begin{eqnarray}
|d_e|< 8.7\times10^{-29} e{\rm cm}\; .
\end{eqnarray}
This limit severely constrains the allowed magnitude of CP-phases in the Higgs couplings
~\cite{Cheung:2014oaa,Inoue:2014nva,Bishara:2013vya,Brod:2013cka} via Barr-Zee diagrams.
We would like to mention that, one must take into account the tension between the CP-phase being required to successfully implement EWBG and constraints from the eEDM.
The preliminary results of the work are listed as follows:
\begin{itemize}[leftmargin=.5cm,rightmargin=.5cm]

\item The EWPT have been explored in two parameter spaces: triple couplings ($c_2,~\delta_1$) and quartic couplings ($d_2,~\delta_2$), as well as cubic and quartic couplings ($\delta_2,c_2$).
The behaviors of the strength of EWPT ($v_c/T_c$) and the energy gap $\Delta V$ ( the energy difference between the Electroweak
vacuum and the vacuum of the complex singlet) are found to be opposite: a smaller $\Delta V$ corresponds to a bigger $v_c/T_c$.
Two types of phase transition, i.e., one- and two-step are studied,
and the two-step phase transition always gives rise to a bigger $v_c/T_c$, which is found to the same as
the real singlet case as explored in~\cite{Profumo:2007wc}.
The dynamics of the phase transition could be characterized by $\Delta V$ to some extent.

\item
The BAU during the EWPT in the model has been explored. The behaviors of BAU as functions of triple couplings and quartic couplings match well with
that of EWPT. The magnitude of BAU is proportional to the strength of the
strong first order electroweak phase transition (SFOEWPT) for the one-step EWPT, and could be a little higher in the two-step EWPT case with the sign-flip behavior appears at some larger $v_c/T_c$.

\item
The imaginary part of the complex singlet serves as the DM candidate. The smaller dark matter mass ( $m_A$) region is severely
constrained by the direct detection experiment LUX.  A larger $m_A$ can give rise to a relatively higher magnitude of the relic density.
The EWPT in the higher magnitude of the dark matter mass region could be independent of $m_A$. A benchmark scenario which accommodates DM and EWBG
is presented. In the scenario, the magnitude of the BAU could fall into the observed range of PLANCK~\cite{Ade:2013zuv} in the two-step SFOEWPT situation, and the CP violation
phase needed for the EWBG is allowed by the constraints of ACME.

\end{itemize}

\section{The model}\label{sec:cxsm_model}
With the notation as in~\cite{Barger:2008jx}, the tree-level potential of the model is given by,
\begin{eqnarray}
\label{eq:potential_z2u1}
 V\of{H,\bigS} &=& \frac{1}{2}m^2H^\dagger H + \frac{\lambda}{4}\of{H^\dagger H}^2+ \frac{\delta_2}{2}H^\dagger H\abs{\bigS}^2
 \nonumber\\
 &&+\Big(\frac{\delta_1e^{i\phi_{\delta_1}}}{4}H^\dagger H\bigS+ \mathrm{c.c.}\Big)
+ \frac{b_2}{2}\abs{\bigS}^2+\frac{d_2}{4}\abs{\bigS}^4\nonumber\\
&&
+\Big(\frac{1}{4}b_1e^{i\phi_{b1}}\bigS^2 +\frac{c_2e^{i\phi_{c2}}}{6}\bigS\abs{\bigS}^2 + \mathrm{c.c.}\Big)\; ,
\end{eqnarray}
where $H$ and $\bigS$ are the $SU(2)$ doublet and complex singlet fields.
For simplicity, phases are chosen as: $\phi_{b_1}=0$ ($\pi$), $\phi_{\delta_1}=0$ ($\pi$), and $\phi_{c_2}=0$ ($\pi$).
To get the minimization conditions of the potential, it is convenient to represent the $SU(2)$ doublet and complex singlet
as $H =(0,~ h/\sqrt{2})$ and $\bigS=(S+i A)/\sqrt{2}$. Thus
Eq.~(\ref{eq:potential_z2u1}) recasts the form of,
\begin{eqnarray}
\label{eq:potential_tree}
 V_0\of{h,S,A} &=& \frac{m^2}{4}h^2 + \frac{\lambda}{16}h^4
+\frac{\sqrt{2}}{8}\delta_1h^2S + \frac{\delta_2}{8}h^2\of{S^2+A^2}\nonumber\\
 &&+ \frac{1}{4}\of{b_2+b_1}A^2
+ \frac{1}{4}\of{b_2-b_1}S^2+ \frac{\sqrt{2}}{12}c_2S\nonumber\\
&&
\times\of{S^2+A^2}+ \frac{d_2}{8}S^2A^2
+ \frac{d_2}{16}\of{S^4 + A^4}\; ,
\end{eqnarray}
in which $\delta_1$ and $c_2$ can be both positive or negative.
As could be seen from Eq.~(\ref{eq:potential_tree}), the breaking of the global $U(1)$ induces one $Z_2$ symmetry for the imaginary part of $\bigS$, i.e. $A$, thus makes a feasible DM candidate. Meanwhile, the cubic and quartic interactions that are analogous to \cite{Profumo:2007wc} are both kept for the EWPT study.

\begin{figure}[!htb]
\centering
	\includegraphics[width=.4\textwidth]{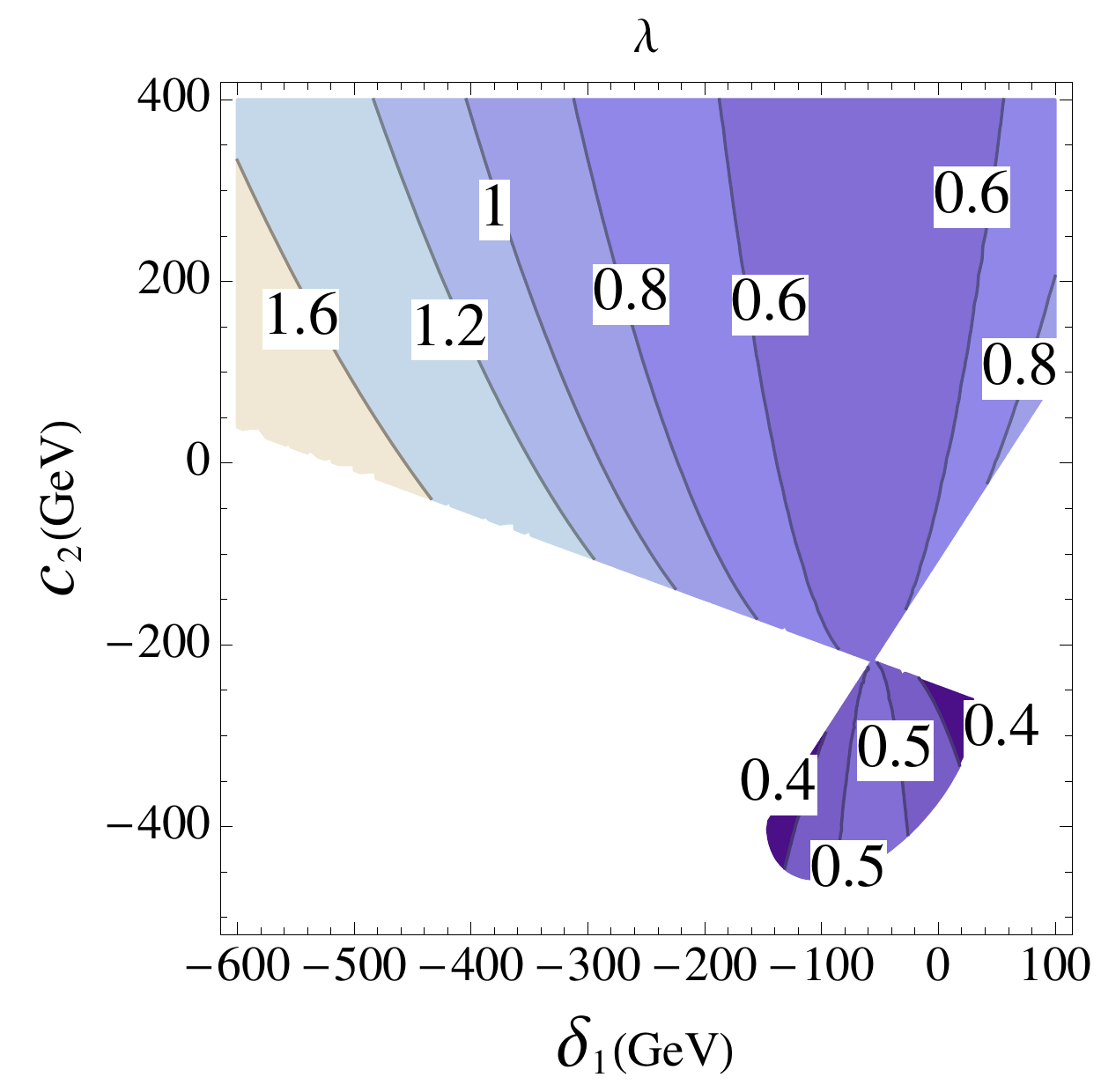} \\
	\includegraphics[width=.4\textwidth]{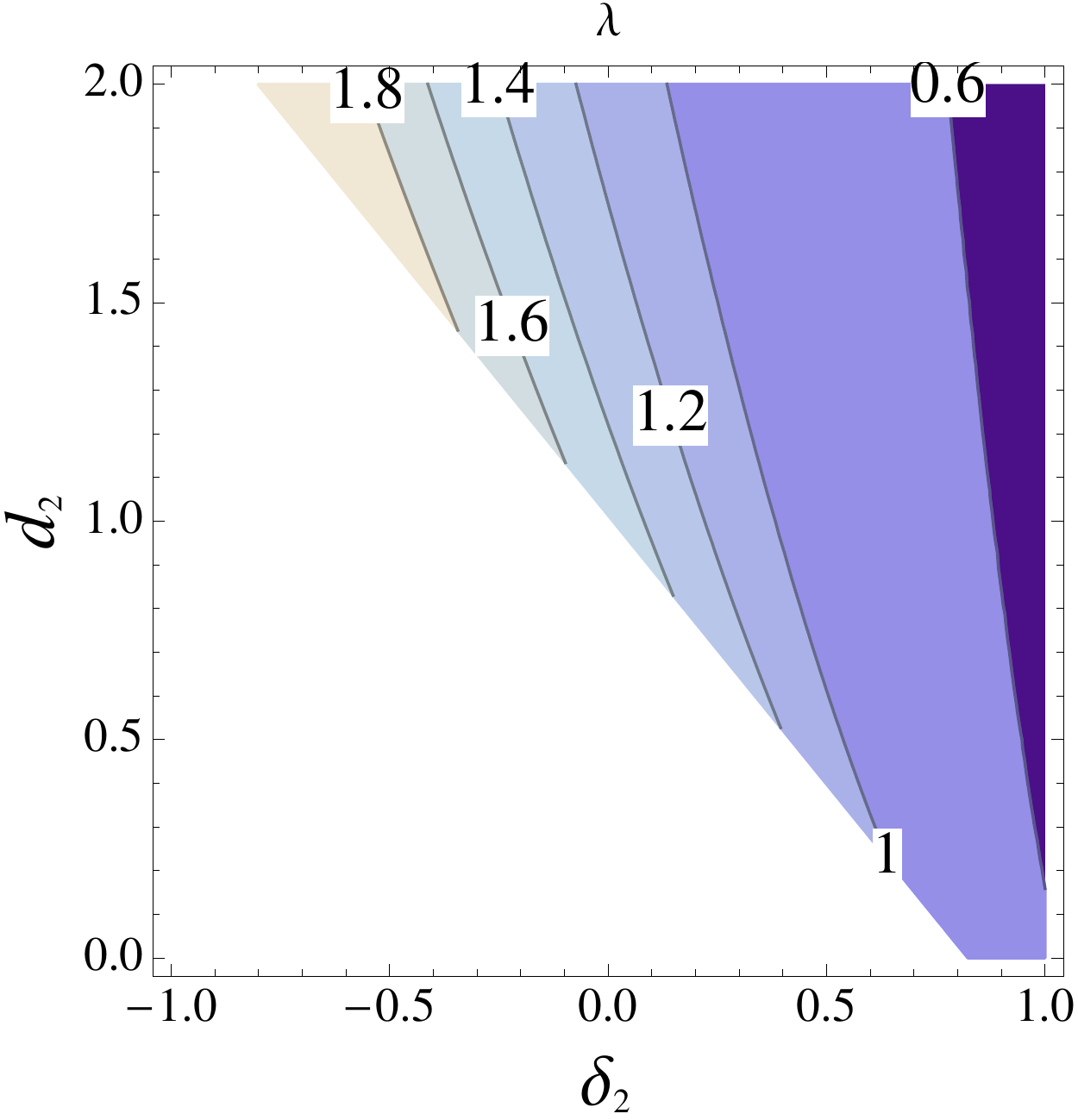}
     \caption{Top: $\lambda$ as a function of $\delta_1$ and $c_2$ from $m_{h_1}=125$ GeV with relevant parameters being set as: $x=200$ GeV,~$m_A =300$~GeV,~$d_2=1.4,~\delta_2=0.2$; Bottom: $\lambda$ as a function of $\delta_2$ and $d_2$ with $m_{h_1}=125$GeV, relevant parameters are set as: $x=200$ GeV,~$m_A =300$~GeV,~$\delta_1=-350$ GeV, and $c_2=100$ GeV.}
\label{fig:mass2}
\end{figure}

The three minimization conditions of the potential,
\begin{eqnarray}
\label{minc}
&&(\partial V_0/\partial h)|_{h=v,S=x,A=0}= 0,\nonumber\\
 &&(\partial V_0/\partial S)|_{h=v,S=x,A=0} =0,\nonumber\\
 &&(\partial V_0/\partial A)|_{h=v,S=x,A=0} =0,
 \end{eqnarray}
 with $v=246.2$ GeV and
$x$ being the vacuum expectation values (VEVs) of the $h$ and $S$,
imply that,
\begin{equation}
\begin{aligned}\label{replace m2,b2}
&m^2 = -\frac{\sqrt{2} \delta_1}{2}x-\frac{\delta_2}{2}x^2-\frac{\lambda}{2}v^2\; ,\\
&b_2=b_1-\frac{\sqrt{2}\delta_1}{4}v^2/x-\frac{\sqrt{2}c_2}{2}x-\frac{d_2}{2}x^2-\frac{\delta_2}{2}v^2\; .
\end{aligned}
\end{equation}
At the minima, the mass matrix of $h$ and $S$ is,
\begin{multline}\label{eq:mat}
\mathbb{M}=\lb\begin{array}{cc}\frac{1}{2}\lambda v^2 & \frac{1}{2}\delta_2 xv+\frac{\sqrt{2}}{4}v\delta_1\\
\frac{1}{2}\delta_2 xv+\frac{\sqrt{2}}{4}v\delta_1 & \frac{1}{2}d_2x^2+\frac{\sqrt{2}}{4}c_2x-\frac{\sqrt{2}\delta_1v^2}{8x}\end{array}\rb \; ,
\end{multline}
and the DM candidate $A$, with mass $m_A^2=b_1 -\frac{\sqrt{2}}{12}c_2x-\frac{\sqrt{2}}{8}\delta_1v^2/x $, does not mix with $h$ or $S$.
The non-zero $m_{hS}^2$ induces mixing between the SM Higgs and the real component of the singlet ($S$).
The mass square of the ``Higgs-like" particle $m_{h_1}^2$ is given by:
 $\frac{1}{2}(m_h^2+m_s^2+\sqrt{(m_h^2-m_s^2)^2+4(m_{hs}^2)^2})$ for $\mbox{$ m_h^2>m_S^2 $}$, and
    $\frac{1}{2}(m_h^2+m_s^2-\sqrt{(m_h^2-m_s^2)^2+4(m_{hs}^2)^2})$ for $\mbox{$m_h^2<m_S^2$}\;$.
    For $m_h^2>m_s^2$, the ``Higgs-like" eigenstate $h_1$ will be the heavier one,
and the ``Singlet-like" eigenstate $h_2$ will be heavier
at the time of $m_h^2<m_s^2$.
Demanding $m_{h_1}=125$ GeV, $\lambda$ could be given as a function of $(c_2,\delta_1)$ or $(d_2,\delta_2)$,
as shown in Fig.~\ref{fig:mass2} .
The mass eigenstates $h_1, h_2$ are related with gauge eigenstates $h$ and $S$ as,
\begin{equation}\label{eq:tree_mass_estates}
\lb\begin{array}{c}h_1 \\ h_2\end{array}\rb = \lb\begin{array}{cc} \cos\phi & \sin\phi \\ -\sin\phi & \cos\phi \end{array}\rb\lb\begin{array}c h \\ S\end{array}\rb\; .
\end{equation}
The mass eigenstates $h_1, h_2$ couple to the fermions and gauge bosons via SM Higgs couplings reduced by a factor of $\cos\phi$ and $-\sin\phi$ respectively.
We take the mixing angle to be $-\pi/4\leq\phi\leq\pi/4$ so that $h_1$($h_2$) is always the ``Higgs-like'' (``singlet-like'') eigenstate.
Using this convention, the mass matrix elements of Eq.~(\ref{eq:mat}) could be expressed as,
\begin{eqnarray}
\label{eq:mass_matrix_transform}
m_h^2=\cos\phi^2m_{h_1}^2+\sin\phi^2m_{h_2}^2\; ,\\
m_s^2=\sin\phi^2m_{h_1}^2+\cos\phi^2m_{h_2}^2\; ,\\
m_{hS}^2=\cos\phi\sin\phi(m_{h_1}^2-m_{h_2}^2)\; ,
\end{eqnarray}
which allows us to replace $\lambda$, $d_2$ and $\delta_1$ with $m_{h_1}$,$m_{h_2}$ and $\phi$. As for the vacuum stability condition, parameters living in $d_2<0$ and $\delta_2<-\sqrt{\lambda d_2}$~\cite{Barger:2008jx,Gonderinger:2012rd} will be excluded. For more studies on the stability problem in this kind of model, we refer to \cite{Costa:2014qga,Coimbra:2013qq,Gonderinger:2012rd}.

\section{EWBG analysis}

Before the EWBG is triggered,
the early Universe is assumed to be hot and radiation-dominated, containing
zero net baryon charge, and the $SU(2)_L\times U(1)_Y$ electroweak symmetry is manifest~\cite{Kirzhnits:1972iw,Kirzhnits:1972ut,Dolan:1973qd,Weinberg:1974hy}.
With the universe cooling to temperature below the electroweak scale, the Higgs field acquires a
non-zero expectation value and thus spontaneously breaks the $SU(2)_L\times U(1)_Y$ symmetry
down to $U(1)_{EM}$. During this phase transition process, the EWBG takes place.

Bubble (of the broken phase) formation and growth begin at nucleation temperature $T_N$( $< T_C$, the
critical temperature of SFOEWPT).
Baryon creation in EWBG takes place in the front (and vicinity) of the expanding bubble walls~\cite{Cohen:1993nk}: With CP violation implemented, particles in the plasma (symmetric phase) scatter with the bubble walls and generate CP (and C) asymmetries in particle number densities;
and these asymmetries diffuse into the symmetric phase ahead of the bubble wall and they bias electroweak sphaleron transitions~\cite{Konstandin:2003dx} to produce more baryons than antibaryons;
and some of the net baryon charge being created is then captured by the expanding wall into the broken phase;
at last, inside the bubbles(broken phase), the rate of sphaleron transitions, which provides (B + L)-violating processes and could wash out the baryons  created, should be strongly suppressed, i.e., the SFOEWPT is needed.

\subsection{SFOEWPT realization}
\label{sfoewpt}
The dynamic of the EWPT is governed by the effective potential at finite temperature, i.e., $V_{eff}\of{h,S,A,T}$ in our model.
After minimizing $V_{eff}\of{h,S,A,T}$, the VEVs
of three scalar fields $h,~S$, and $A$ at each temperature could be obtained.
When a first order phase transition takes place, a local minimum with $\langle h \rangle\neq 0$ develops and becomes degenerate with the electroweak symmetric minimum as the temperature decreases to the critical temperature $T_c$, and the two minima are separated by an energy barrier. To protect the
BAU generated by EWBG from being washed out,
the
strong first order condition \cite{Moore:1998swa},
\begin{equation}
\frac{v_c}{T_c}>1\; ,
\end{equation}
with $v_c$ being the Higgs VEV of the symmetry broken minimum, is necessary.
Here, when one uses the condition, the following three most important theoretical uncertainties need to be kept in mind \cite{Patel:2011th}: $v_c$ in the condition is indeed the value of the classic field at minima at $T_c$($\phi(T_c)$) which depends on the gauge choice; the gauge dependence of $T_c$ does not compensate that of $\phi(T_c)$; and the choice of the washout factor is subject to additional uncertainties.

The standard analysis of $V_{eff}\of{h,S,A,T}$ involves
tree level scalar potential ($V_0\of{h,S,A}$) and Coleman-Weinberg one-loop effective potential ($V_1\of{h,S,A}$)
at $T=0$,
the counter terms  $V_{ct}$ being chosen to maintain the tree level relations of the parameters in
the $V_0\of{h,S,A}$, and the leading thermal corrections being denoted by $V\of{h,S,A,T}$,
\begin{align}
V_{eff}\of{h,S,A,T} &= V_0\of{h,S,A}+V_1\of{h,S,A}+ V\of{h,S,A,T}\nonumber\\
&\quad+V_{ct}\of{h,S,A}\; .
\end{align}
Here, $V_0$ is the same as the tree level potential Eq.~(\ref{eq:potential_tree}), $V_1$ is the Coleman-Weinberg one-loop effective potential at $T=0$ and could be expressed in terms of the field-dependent masses $m_i\of{h,s,A}$:
\begin{equation}
V_1\of{h,S,A} = \sum_{i}n_i\frac{m_i^4\of{h,S,A}}{64\pi^2}\left[\ln\of{\frac{m_i^2\of{h,S,A}}{Q^2}}-C_i\right] \;
\end{equation}
in $\overline{MS}$ scheme and Landau gauge.
Here, the sum $i$ runs over the scalar, fermion, and vector boson contributions. $n_i$ is the number of degrees of freedom for the $i$-th particle, with a minus sign for fermionic particles. $Q$ is a renormalization scale which we fix to $Q=v$. We take
$C_i=\frac{1}{2}$ ($C_i=\frac{3}{2}$) for the transverse (longitudinal) polarizations of gauge bosons, and $C_i=\frac{3}{2}$ for all other particles. The field-dependent masses are given in the Appendix.~B.
With $V_1$ being entailed in the potential, the minimization conditions Eq.~(\ref{minc}) will be shifted slightly, and the relations Eq.~(\ref{replace m2,b2}) do not hold as well. To maintain
these relations, the counter terms $V_{ct}$ should be added~\cite{Fromme:2006cm,Cline:2011mm}, and cast the form of,
\begin{equation}
V_{ct}=\delta m_1^2 h^2+\delta m_2^2 S^2+\delta m_3^2 A^2\; ,
\end{equation}
in which, the relevant coefficients are determined by,
\begin{eqnarray}
\delta m_1^2=-\frac{1}{2v}\frac{\partial V_1}{\partial h}|_{h=v,S=x,A=0}\; ,\\
\delta m_2^2=-\frac{1}{2x}\frac{\partial V_1}{\partial S}|_{h=v,S=x,A=0}\; ,\\
\delta m_3^2=-\frac{1}{2}\frac{\partial^2 V_1}{\partial A^2}|_{h=v,S=x,A=0}\; .
\end{eqnarray}
Thus, the VEVs of $h$ and $S$ as well as the dark matter mass will not be shifted. We do not include more
complicate terms to compensate the shift of mass matrix of $S$ and $h$, because
 these shift effects are basically small.

The finite temperature component $ V\of{h,S,A,T}$ is obtained as
\begin{eqnarray}
 V\of{h,S,A,T} = \frac{T^4}{2\pi^2}\, \sum_i n_i J_{B,F}\lp\frac{ M_i^2(h,A,S,T)}{T^2}\rp\;,
\end{eqnarray}
where the functions $J_{B,F}(y^2)$ are given by
\begin{eqnarray}
J_{B,F}(y) = \int_0^\infty\, dx\, x^2\, \ln\left[1\mp {\rm exp}\left(-\sqrt{x^2+y}\right)\right]\; ,
\end{eqnarray}
with the upper (lower) sign corresponds to bosonic (fermionic) contributions. And $M_k^2(h,S,A,T)$ is given in terms of the field dependent mass at $T=0$ and the finite temperature mass function $\Pi_k$ as,
\begin{eqnarray}
 M_k^2(h,S,A,T)=m_k^2(h,S,A)+\Pi_k\ \;, \ \
\end{eqnarray}
with $\Pi_k$ being listed in Appendix.~B.

\begin{figure}[!htp]
\centering
\includegraphics[width=.4\textwidth]{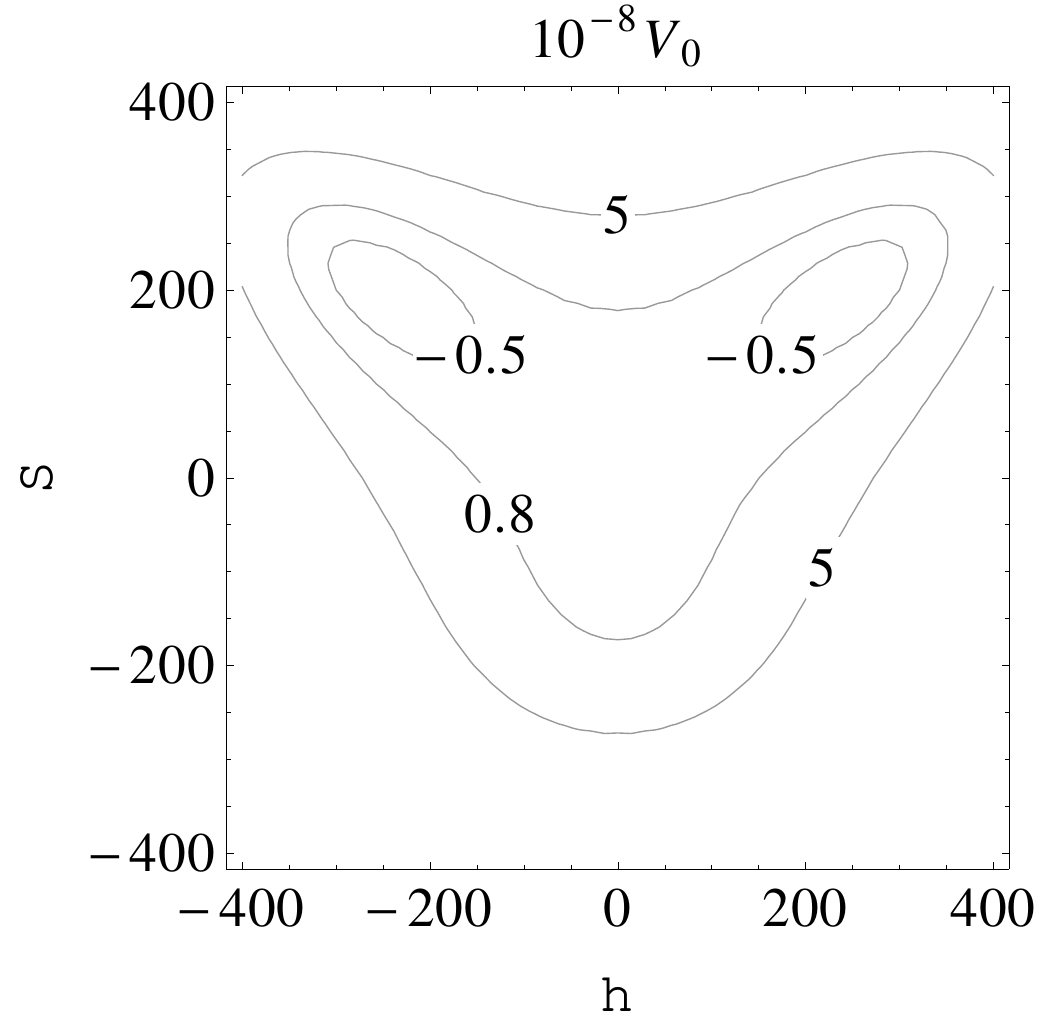}
\includegraphics[width=.4\textwidth]{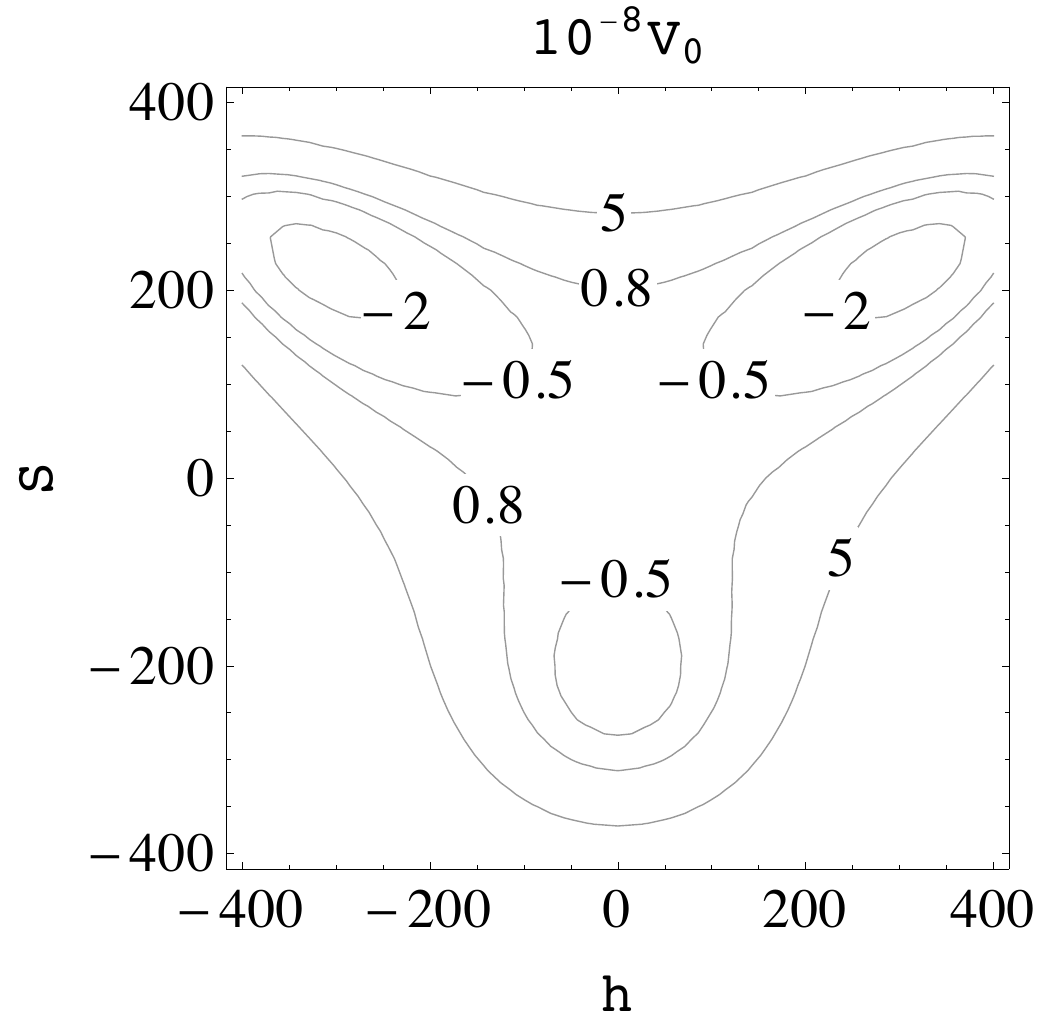}
\caption{The contours of $V_0$ (in units of GeV$^4$) in $h-S$ ( in units of GeV ) plane. With parameters being fixed as: $x=200$ GeV, $\delta_1$=-300 GeV, d2=1.4, $\delta_2$=0.2, $\lambda$=1.0,  and $c_2=-10$ ($100$) GeV for the top (bottom) panel.}
\label{fig:v0}
\end{figure}

The behaviour of EWPT strongly depends on the shape of the potential at zero temperature.
Fig.~\ref{fig:v0} shows contours of $V_0$ in the $h-S$ plane in two typical situations.
Since the $V_0$ does not develop a minimum at $A\neq 0$ when the dark matter mass is large as will be explored in Sec.~\ref{dmewpt}, we conduct the analysis in the $A=0$ plane.
In the top panel, the electroweak vacuum is the only local minimum of the potential $V_0$. For this situation, the universe might transform from the high-T  symmetric phase to the electroweak symmetry broken phase directly.
In the bottom panel,
an additional local minimum besides the electroweak vacuum appears in the direction of $S$ at $h=0$. For this situation, there is a possibility that the universe transforms to this phase at the first step and then to the electroweak symmetry broken phase at the second step.
And in fact, the first investigation of the real singlet case~\cite{Profumo:2007wc} shows that the two-step transition
(with the second step from the high temperature singlet vacum to the electroweak vacuum) gives rise to a higher magnitude of EWPT strength in comparison with the one-step transition case.~\footnote{It is also feasible to study fermion dark matter together with strong first order
phase transition~\cite{Li:2014wia}.} Since the potential at this additional minimum (in the direction of $S$) can be very close to that of the electroweak vacuum, the second step can happen at a very low temperature, therefore a strong first order phase transition could be realized easily. Of course, the phase transition could also be
the one-step case, as long as the energy gap between the minima in the direction of $S$ and $h$ is too large to be overcome by the T dependent
thermal correction at high temperature. For the similar analysis in a triplet scenario and Next-to-Minimal Supersymmetric Standard Model (NMSSM) we refer to~\cite{Patel:2012pi} and \cite{Huang:2014ifa,Kozaczuk:2014kva,Bi:2015qva}.

\begin{figure}[!htb]
\centering
\includegraphics[width=.4\textwidth]{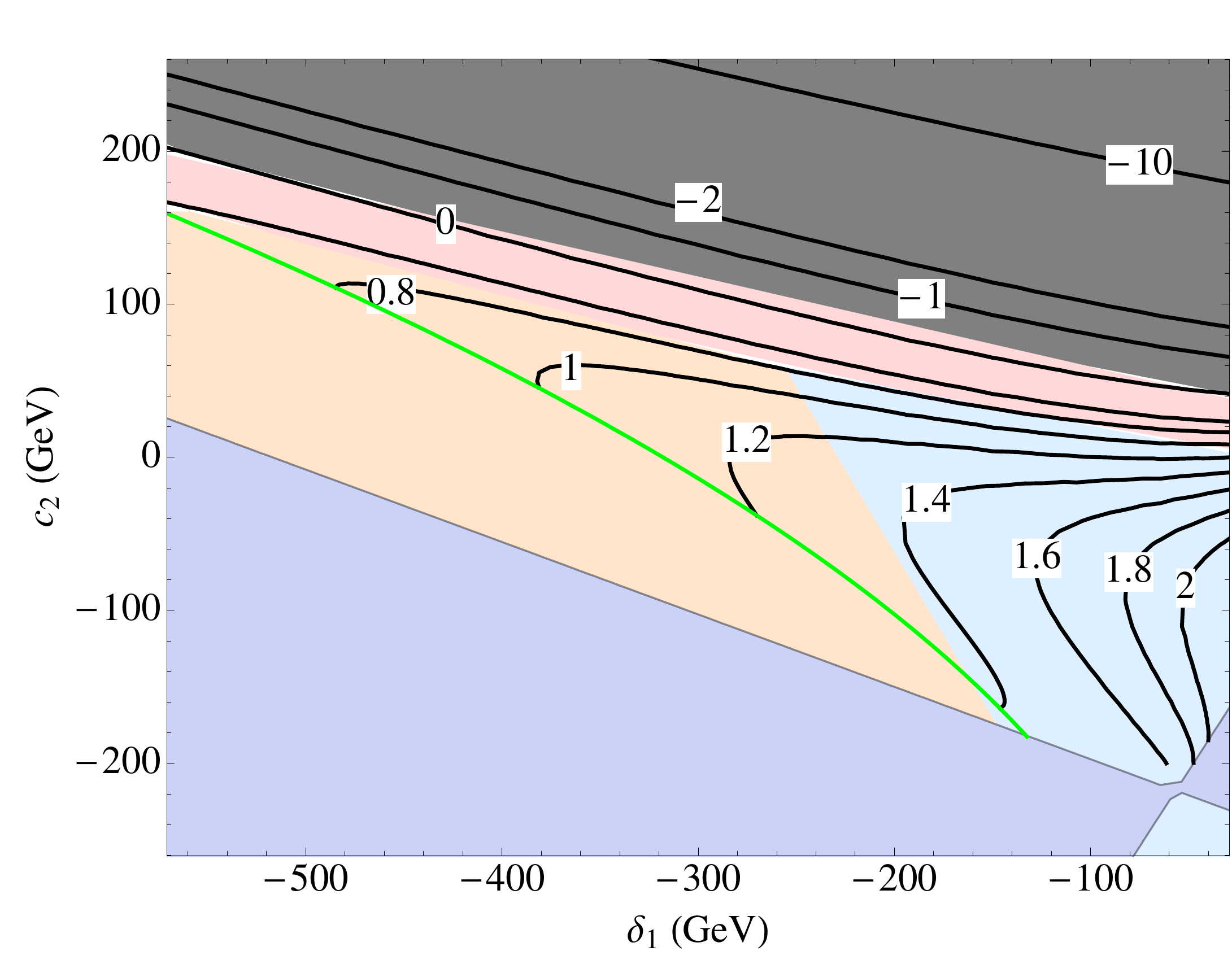}\\
\includegraphics[width=.4\textwidth]{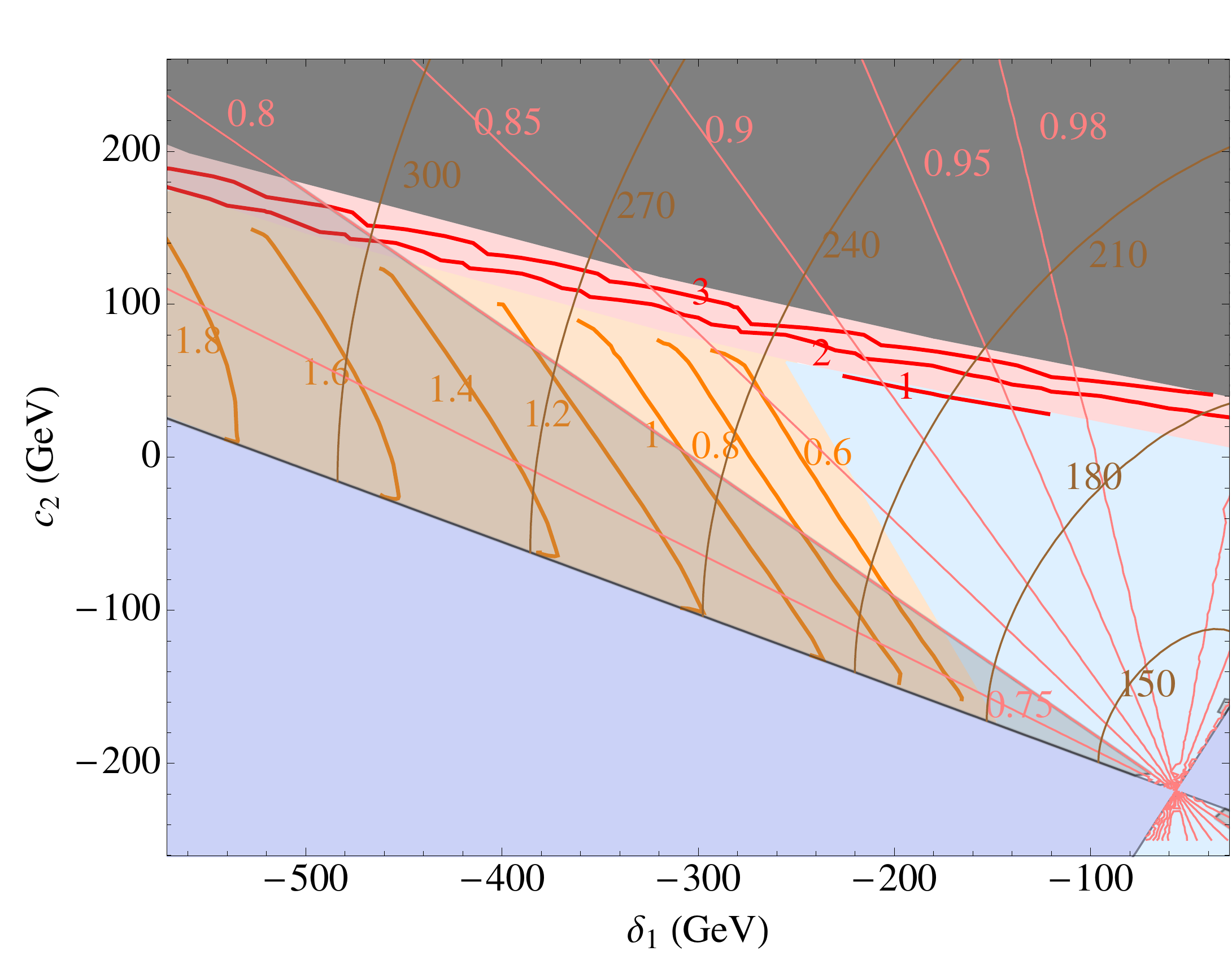}
\caption{EWPT with parameters being fixed as: $x =200$~GeV,~$m_A =300$~GeV, $\delta_2 =0.2$,~$d_2 =1.4$, and $\lambda$ being solved from $m_{h_1}=125$ GeV as shown in Fig.~\ref{fig:mass2}. The blue shade is the regions where $m_{h1}=125$ GeV couldn't be obtained. The grey region depicts that the electroweak vacuum is a local but not global minimum. The light orange and red regions are parameters space where one-step and two-step first order phase transitions take place. The second order phase transition happens in the light blue region. Top:
The green line is to divide the parameter spaces into the regions where the minimum in the direction of $S$ exists (above) and disappears (below). The black contours in the region above the green line are the contours of $\Delta V=V_S-V_h$ in units of $10^8$ GeV$^4$; Bottom: The contours of $v_c/T_c$, $\cos\phi$, and $m_{h_2}$ are shown with orange(and red), pink and brown lines. The region excluded by $\cos\phi<0.8$, which characterizes the mixing between $h$ and $S$, is covered by light grey shade.}
\label{fig:c2-delta1}
\end{figure}

We take advantage of the {\tt CosmoTransitions} package~\cite{Wainwright:2011kj} to investigate the phase transition numerically. Fig.~\ref{fig:c2-delta1} shows the result in the $c_2-\delta_1$ plane. In the top panel,
the contour lines of $V_0$ are plotted to help understand this pattern qualitatively: for parameter regions bellow the green line, the electroweak vacuum is the only minimum of $V_0$, thus only one-step phase transition could happen.
 The contour $\Delta V=0.6\times 10^8$~GeV$^4 $ separates the one-step from the two-step type phase transition.
 When the difference between the two minima is bigger than
 that value, the T dependent thermal corrections could only changes the magnitude of $\Delta V$
 a bit thus only the one-step phase transition could be obtained. When the magnitude of $\Delta V$
 gets smaller, the phase transition type changes from one-step to two-step.
The bottom panel indicates that, with the increase of $c_2$ and $\delta_1$, one gets smaller $v_c/T_c$  for the one-step phase transition, and a second order phase transition is obtained at last. For the two-step cases, $v_c/T_c$ gets larger with the increase of $c_2$ and $\delta_1$ and the SFOEWPT condition $v_c/T_c >1$ could always be satisfied. The pattern of the contours of $v_c/T_c$ is opposite to that of $\Delta V$, i.e., the smaller $\Delta V$ leads to the larger $v_c/T_c$.

\begin{figure}[!ht]
\centering
\includegraphics[width=.4\textwidth]{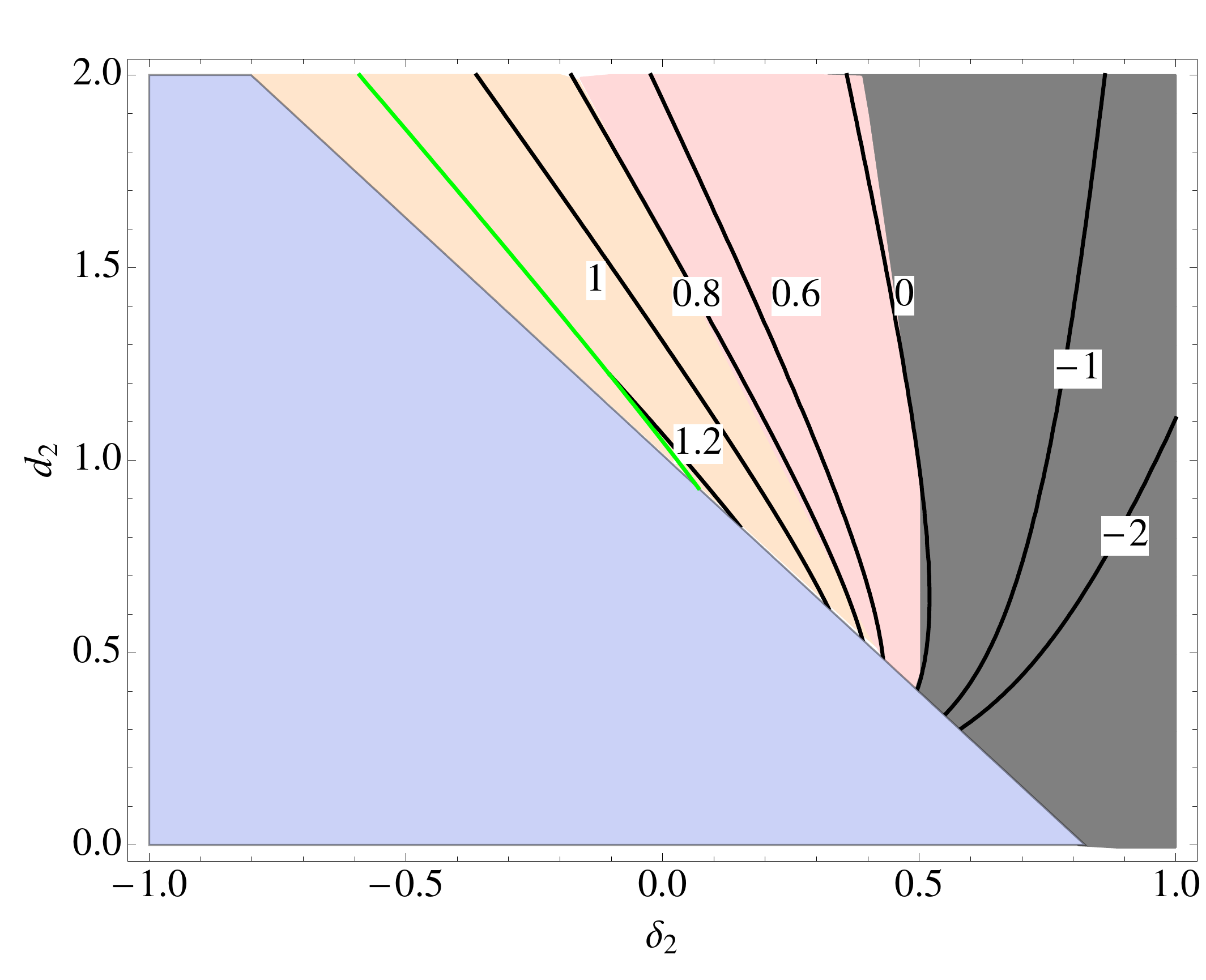}\\
\includegraphics[width=.4\textwidth]{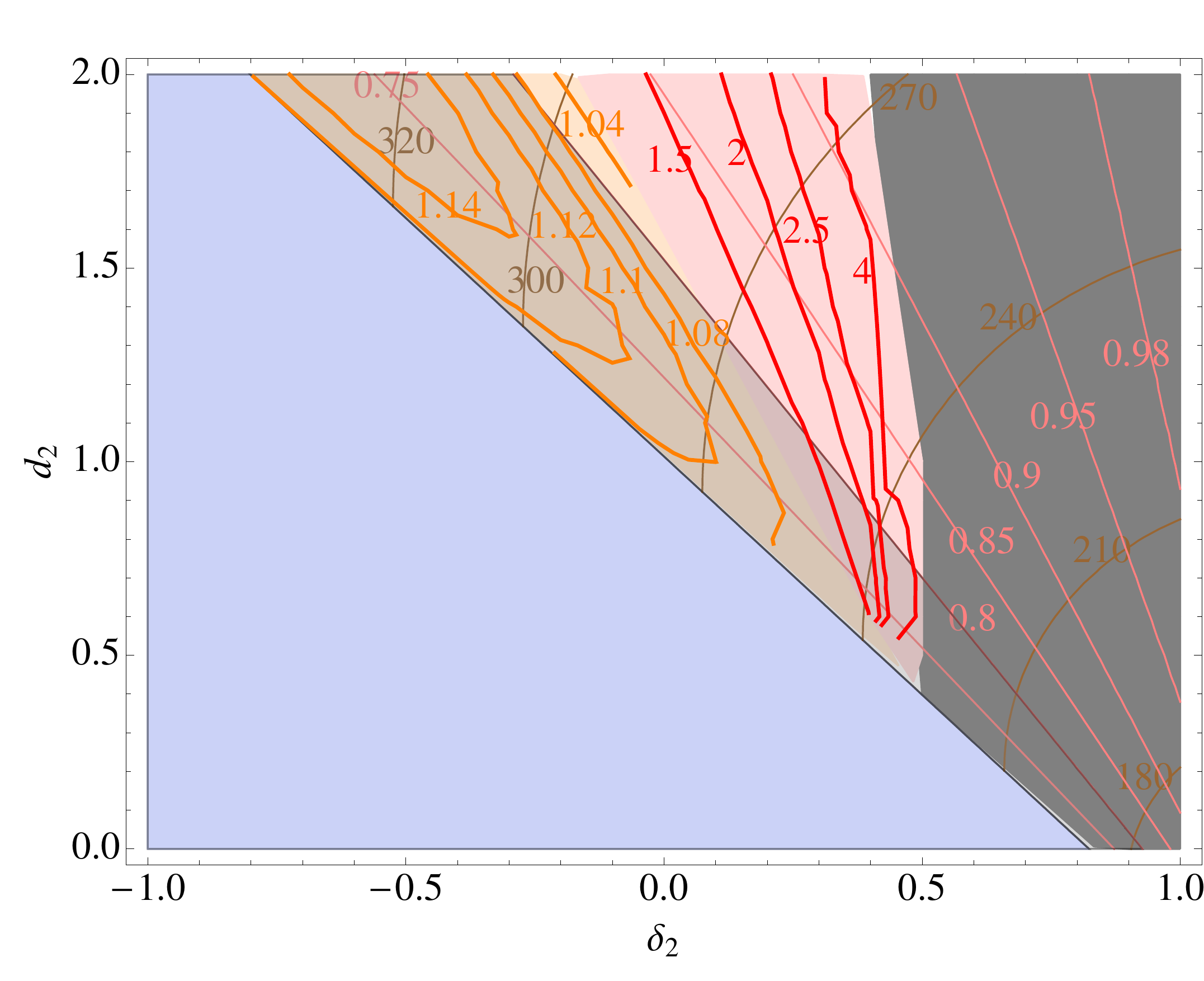}
\caption{ With parameters being fixed as $x=200$~GeV,~$m_A =300$~GeV,~$\delta_1=-350$ GeV, and $c_2=100$ GeV.
The descriptions of the two pictures are the same as in Fig.~\ref{fig:c2-delta1}.}
\label{fig:delta2-d2}
\end{figure}

Fig.~\ref{fig:delta2-d2} illustrates the dependence of the behavior of phase transition on the quartic couplings $\delta_2$ and $d_2$. For the one-step phase transition, the magnitude of $v_c/T_c$ becomes higher and higher with the decrease(increase) of $\delta_2$ ($d_2$). For the two-step type phase transition, the value of $v_c/T_c$ increases with the two couplings getting larger.

As depicted by the bottom panels of Fig.~\ref{fig:c2-delta1} and Fig.~\ref{fig:delta2-d2}, for our parameter choice,
one-step type phase transition parameter regions are severely constrained by the
mixing angle by fitting current Higgs measurements at the LHC~\cite{Profumo:2014opa},  while most two-step ones survive.
And to make a decisive conclusion on this topic, more data are expected from the run II of the LHC as well as other future colliders, such as the International Linear Collider (ILC), TLEP(renamed as FCC-ee recently), China Electron Positron Collider (CEPC), or the Super proton-proton Collider (SPPC), etc. Here, we would like to mention that the electroweak precision observables are supposed to constrain the Higgs and singlet mixing angle $\phi$ and the heavy Higgs mass $m_{h_2}$\cite{Barger:2008jx,Gonderinger:2012rd,Profumo:2007wc,Lopez-Val:2014jva}, which would also bound the SFOEWPT and BAU favored regions.
We plot the contours of the mixing angle and masses of the ``Singlet-like" particle in our parameter regions to take into account this thing. For detailed analysis we refer to~\cite{Profumo:2014opa}.

\subsection{BAU gereratation}
\label{baucal}
\begin{figure}[!htp]
\centering
\includegraphics[width=.4\textwidth]{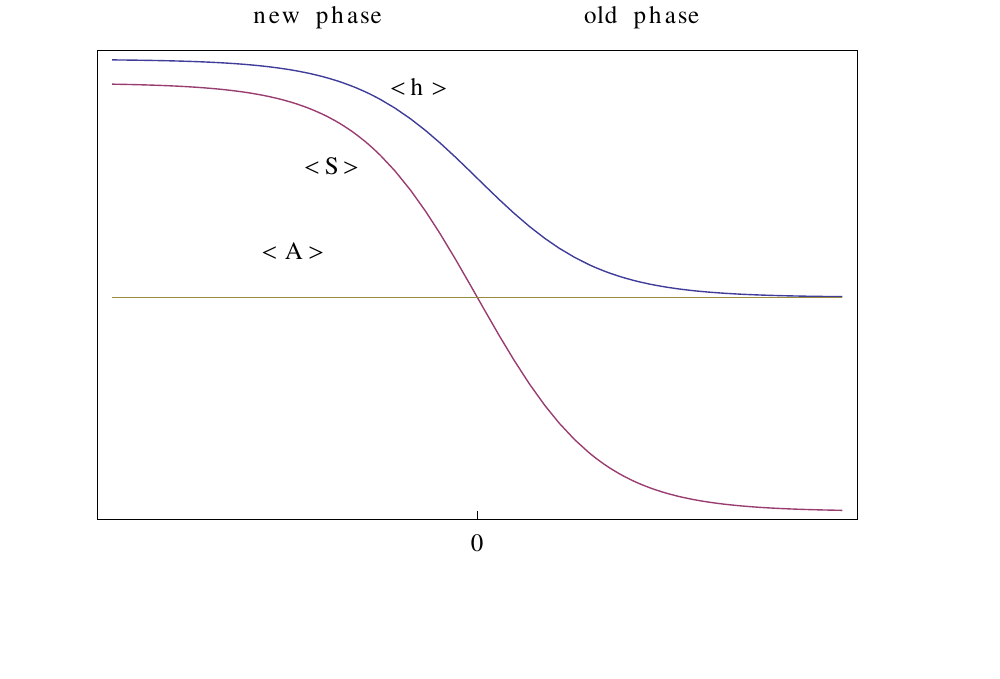}
\caption{The wall profiles at $T_c$ with new (old) phase corresponding to Electroweak broken (unbroken) phase.}
\label{fig:bauprofile}
\end{figure}

To get a more comprehensive understanding of EWBG, we solve the transport equations as explored in Ref.~\cite{Fromme:2006wx} directly,
in which the speed and profile of the walls are the core ingredients.
With the nucleating bubbles reaching a sizable extent and expanding with a constant velocity,
it is appropriate to boost into the rest frame of the bubble wall and proceed all calculations with a planar wall.
With $z$ being the coordinate transverse to the wall,
$h(z)$, $S(z)$ and $A(z)$ are the expectation values of $h$, $S$ and $A$ inside and outside the bubbles.
To determine them precisely, one needs to minimize the Euclidean action~\cite{Espinosa:2011ax}:
\begin{eqnarray}
	S &=& \int_{-\infty}^{\infty}dz\Big[\frac{1}{2}(\partial_zh)^2+\frac{1}{2}(\partial_zS)^2\nonumber\\
	&&+\frac{1}{2}(\partial_zA)^2+V_{eff}(h,S,A,T_c)\Big]\; .
\end{eqnarray}
As a good approximation, we estimate the bubble wall profiles in the form of,
\begin{eqnarray}
  h(z)&=&\frac{1}{2}v_c(1-\tanh(z/L_w))\; ,\\
  S(z)&=&x_c+\frac{1}{2}\Delta x_c(1+\tanh(z/L_w))\; ,\\
  A(z)&=&a_c+\frac{1}{2}\Delta a_c(1+\tanh(z/L_w))\; .
\end{eqnarray}
Here, $L_w$ is the width of the wall, $v_c$, $x_c$, and $a_c$ are the VEVs of the electroweak symmetry broken phase at $T_c$, and $\Delta x_c$ ($\Delta a_c$) is the total change of $\langle S \rangle$ ($\langle A \rangle$). For the scenario of this work, $A$ hardly gets VEV during the whole phase transition process,  therefore, $A(z)$ is zero, as will be explored more in section~\ref{dmewpt}. The wall width is approximated with $L_w\approx\sqrt{\of{\Delta x_c^2+\Delta a_c^2+v_c^2}/8V_x}$~\cite{Bodeker:2004ws},  in which $V_x$ is the height of the potential barrier at $T_c$.  The bubble wall profiles are shown in Fig.~\ref{fig:bauprofile}.

As mentioned in Sec.~\ref{sec:intro}, we introduced the CP violating top quark mass term,
and for definiteness we fix $a=0$ to maximize the CP violation.
The top quark acquires a
spatially varying complex mass inside the bubble walls during the phase transition, being obtained as 
\begin{equation}
	m_t(z) = {y_t\over\sqrt{2}} h(z) \left(1 + i {S^2(z)+A^2(z)\over\Lambda^2}\right)\equiv
		|m_t(z)| e^{i\theta(z)}\; .
\end{equation}
following the method used in.~\cite{Espinosa:2011eu,Cline:2012hg}.
We will verify if the nontrivial phase $\theta(z)$ is sufficient to source the baryon asymmetry in the following studies.
 The energy scale will be chosen as $\Lambda=1000$ GeV in relevant calculations of this work, i.e., eEDM and BAU.\footnote{Here, we would like to mention that
the energy scale $\Lambda$ characterizes the size of $\theta(z)$ to some extent. And a larger $\Lambda$ always lead to a smaller $\theta(z)$ and results in a lower magnitude of $\eta_B$ and $de/e$. The dimension six operator in Eq.~(\ref{dim6op})
induces negligible effects in the EWPT process, and the wall profiles shown in Fig.~\ref{fig:bauprofile} ensures that the operator's contribution to $m_t(z)$ is small in comparison with the renormalizable one~\cite{Cline:2012hg}.  }

With the top quark profile of our model being obtained,  we solve the transport equations ( see Apendix D for more details) to get the chemical potentials of left-handed SU(2) doublet top
quark( $\mu_{t,2}$),  left-handed  SU(2) doublet bottoms ($\mu_{b,2}$), left-handed SU(2) singlet top quark ($\mu_{t^c,2}$),  and Higgs bosons ($\mu_{h,2}$), and the corresponding plasma velocities.
The chemical potential of left-handed quarks cast the form of,
 \begin{equation}
\mu_{B_L}=\frac{1}{2}(1+4K_{1,t})\mu_{t,2}+\frac{1}{2}(1+4K_{1,b})\mu_{b,2}
-2K_{1,t}\mu_{t^c,2}\; ,
\end{equation}
where $K$ factors are thermal averages~\cite{Fromme:2006wx}.

Finally, The baryon asymmetry is obtained using
\begin{equation} \label{eta1}
\eta_B=\frac{n_B}{s}=\frac{405\Gamma_{ws}}{4\pi^2v_wg_*T}\int_0^{\infty}
dz ~\mu_{B_L}(z)e^{-\nu z}\; ,
\end{equation}
where $\Gamma_{ws}$ is the weak sphaleron rate and
$\nu=45\Gamma_{ws}/(4v_w)$. The effective number of
degrees of freedom in the plasma is $g^\star=106.75$.
And we take the wall velocity to be $v_w=0.1$ in our calculation.

\begin{figure}[!htp]
\centering
\includegraphics[width=.4\textwidth]{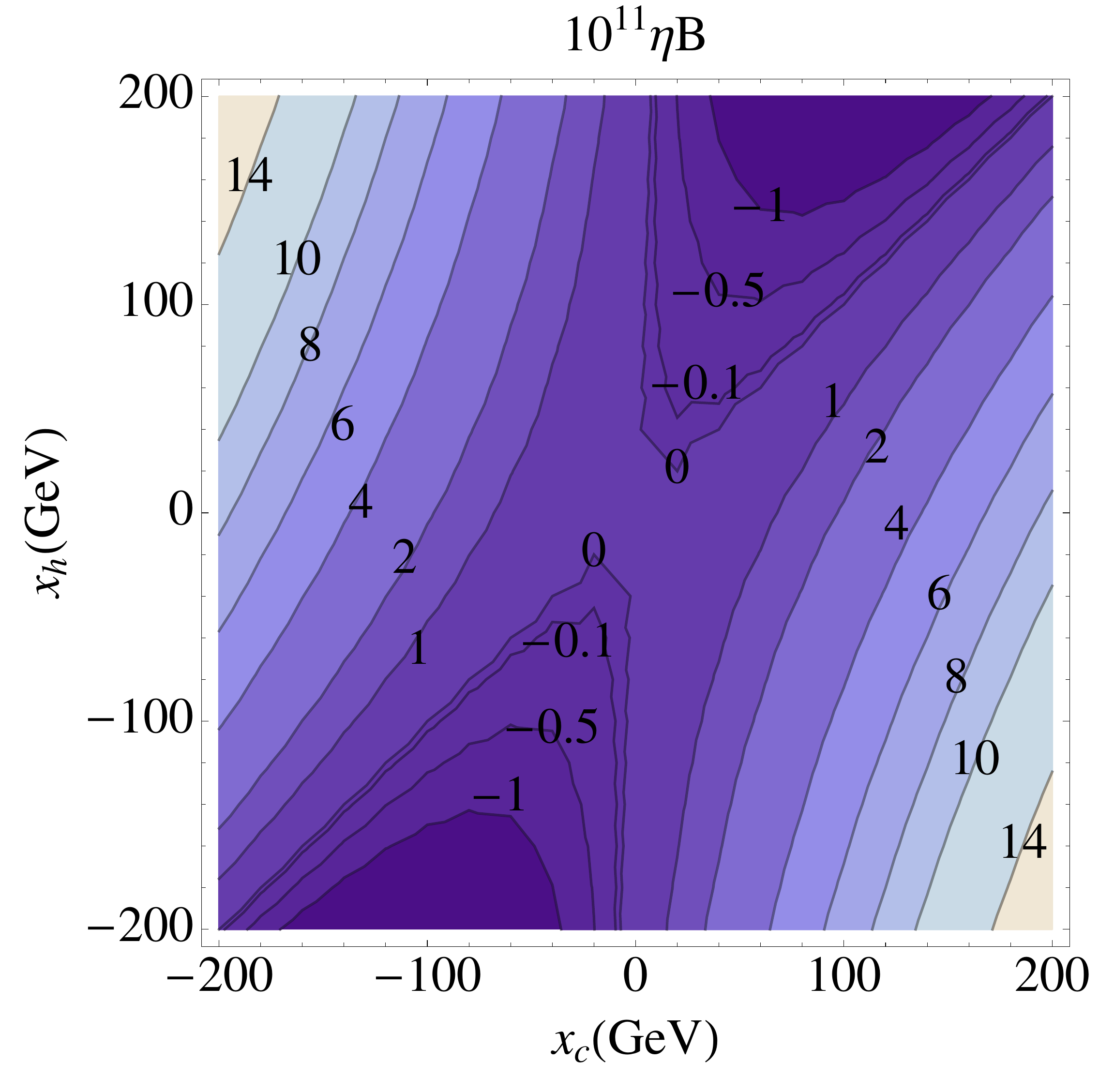}
\includegraphics[width=.4\textwidth]{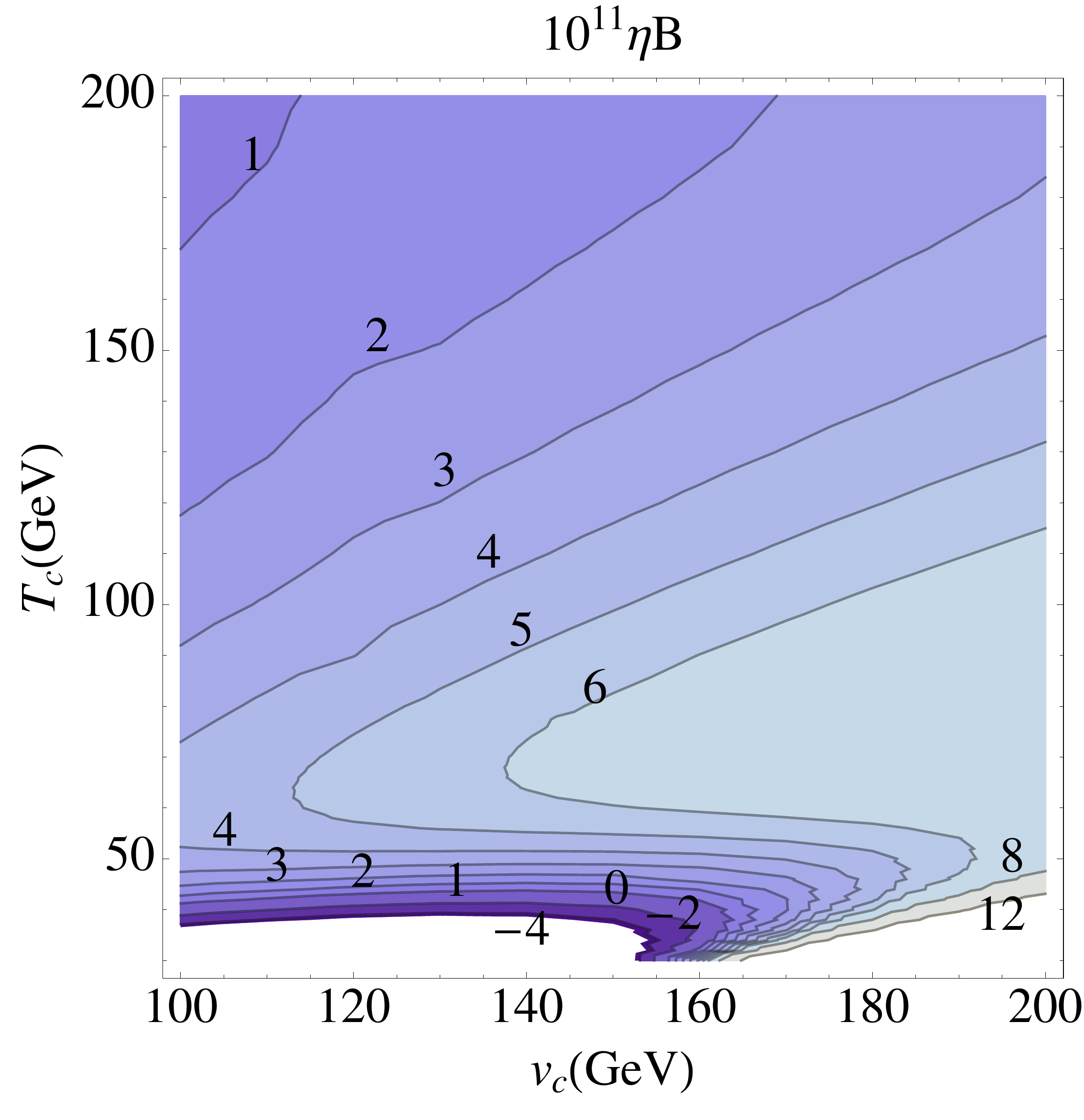}\\
\includegraphics[width=.4\textwidth]{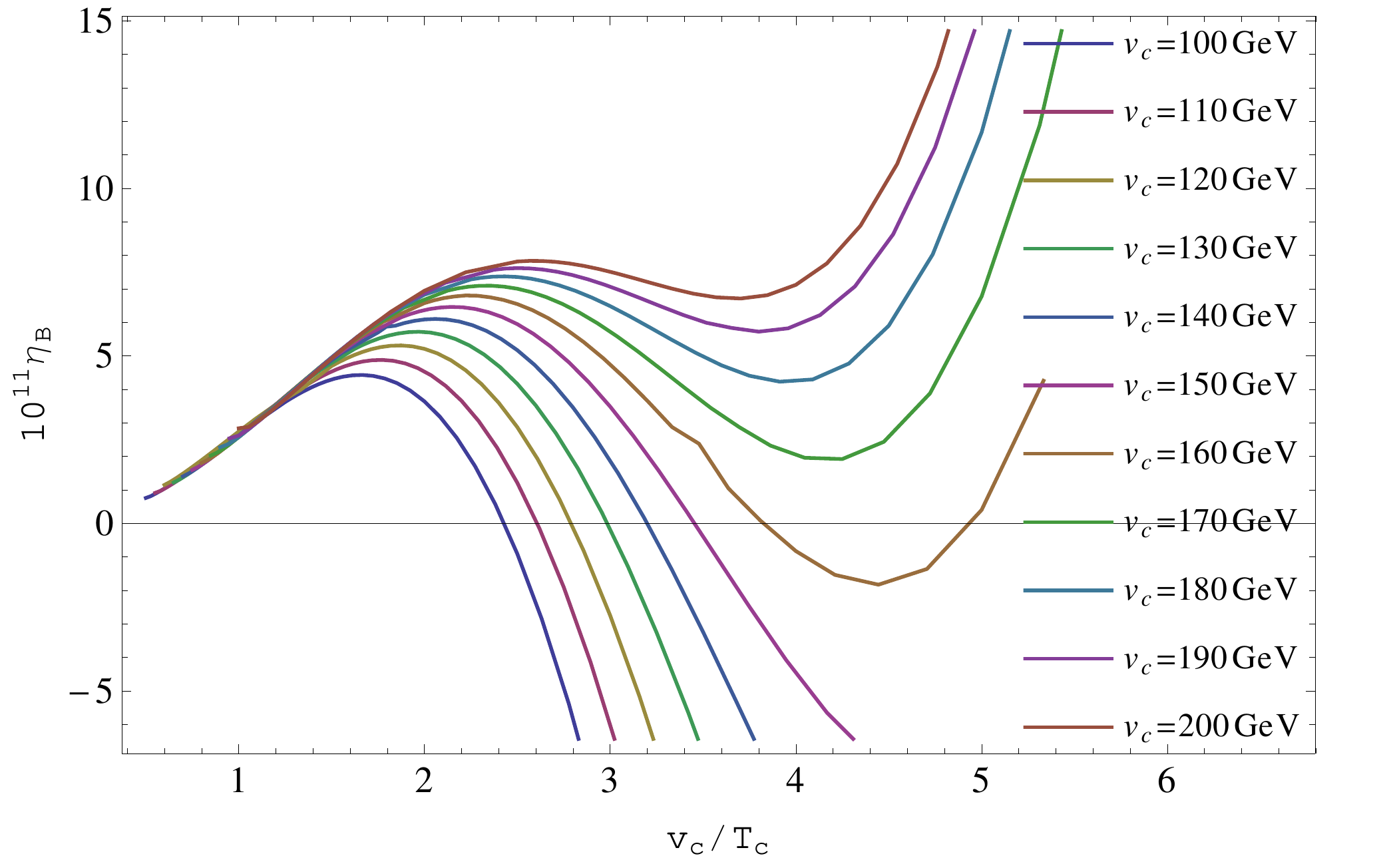}
\caption{The correlation between baryon asymmetry and the EWPT parameters.  Parameters are fixed as
 $v_c=150$ ~GeV, $a_h=0$, and $T_C=100$ ~GeV($x_c=150$~GeV, $x_h=0$, and $a_h=100$ GeV) for the top (middle) panel.  The magnitude of $\eta_B$ as a function of $x_c$ and $x_h$, $v_c$ and $T_c$, and $v_c/T_c$ are shown in top, middle, and bottom panels .  }
\label{fig:bauresearch}
\end{figure}

To get a first glimpse of  the behavior of the baryon asymmetry with respect to the parameters of EWPT ( i.e.,  the critical temperature $T_c$ and the field expectation
values $v_c$,$x_c$,$x_h$...), we perform the calculation of BAU with these parameters being input directly. We need to note
that the values of $v_c$ and $T_c$ used here are not physical parameters we get during EWPT as explored in the previous section.
The top panel in Fig.~\ref{fig:bauresearch} illustrates how the expectation values of $S$ affect the magnitude of BAU, with $x_c$ and $x_h$ denoting the VEV of $S$ at electroweak symmetry broken phase and symmetric phase.
It depicts that $\eta_B$ gets larger with $x_h$ getting smaller for $x_c>0$, and the situation changes to be opposite when $x_c<0$.
$\eta_B$ changes its sign near $x_h=x_c$. In compare with that of $x_h$, $\eta_B$ relies on $x_c$ more, especially when $|x_c|$ is small.
The middle panel indicates how $\eta _B$ depends on $v_c$ and $T_c$.
The dependence of $\eta_B$ on $v_c$ is simple and obvious: a larger value of $\eta_B$ corresponds to a larger $v_c$ as expected,
since the source of CP violation depends partly upon the amount of variation of $|m_t|$($\sim v_c$) inside the bubble wall.
The dependence of $\eta_B$ on $T_c$ is more complicated.
$\eta_B$ have a almost linear correlation with $T_c$ for larger $T_c$, and the sign of $\eta_B$ flips for smaller $T_c$. The bottom panel gives the magnitude of BAU as a function of $v_c/T_c$, which
indicates that some value of $v_c$ might induce unphysical results, i.e., $\eta_B\le 0$. Thus, a larger $v_c/T_c$ may not leads to a bigger $\eta_B$.

\begin{figure}[!htp]
\centering
\includegraphics[width=.4\textwidth]{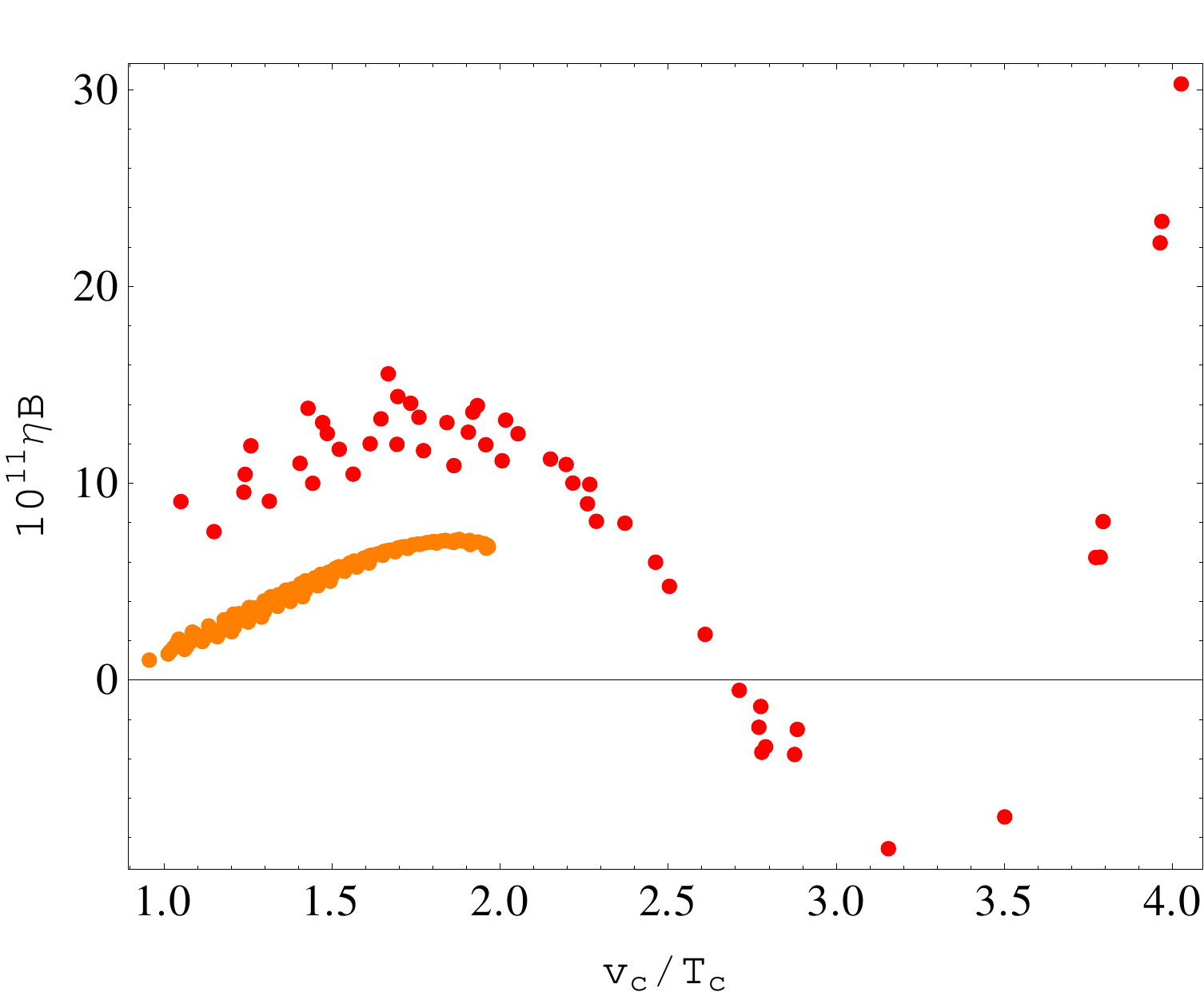}
\includegraphics[width=.4\textwidth]{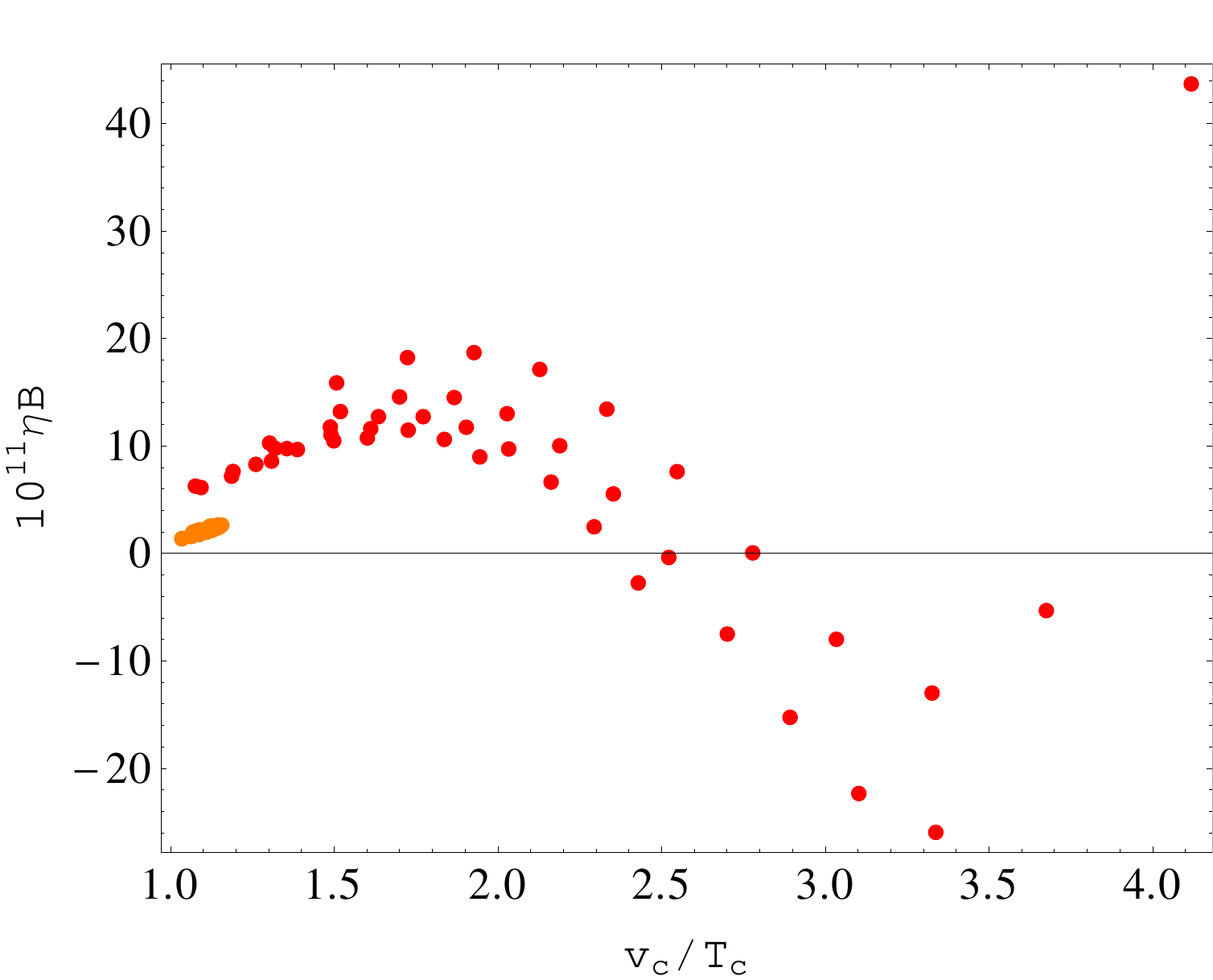}
\caption{The BAU magnitude $\eta_B$
as a function of $v_c/T_c$, with orange(red) points depicting the one-step(two-step) phase transition cases. Parameters are set as: $x=200$~GeV,~$m_A =300$~GeV,~$\delta_2 =0.2$ ( $\delta_1=-350$~GeV) and $d_2 =1.4$ ($c_2=100$~GeV) for top (bottom) panel.
.}
\label{fig:baurecalc2}
\end{figure}

Then, we use the results obtained in the section.~\ref{sfoewpt}
as the input parameters and
calculate the magnitude of the BAU being generated during SFOEWPT.
The results are shown in Fig.~\ref{fig:baurecalc2},
which depicts that for the one-step case the behaviors of $\eta_B$ are very regular and indicates one good correspondence with the behavior of $v_c/T_c$, i.e., a simple linear correlation.
While, a more complicated correlation for the two-step case exists. The sign of $\eta_B$ flips and flips again with the increase of $v_c/T_c$ in the two-step case, which depicts a similar behavior as that of Fig.~\ref{fig:bauresearch}.

\begin{figure}
\includegraphics[width=.45\textwidth]{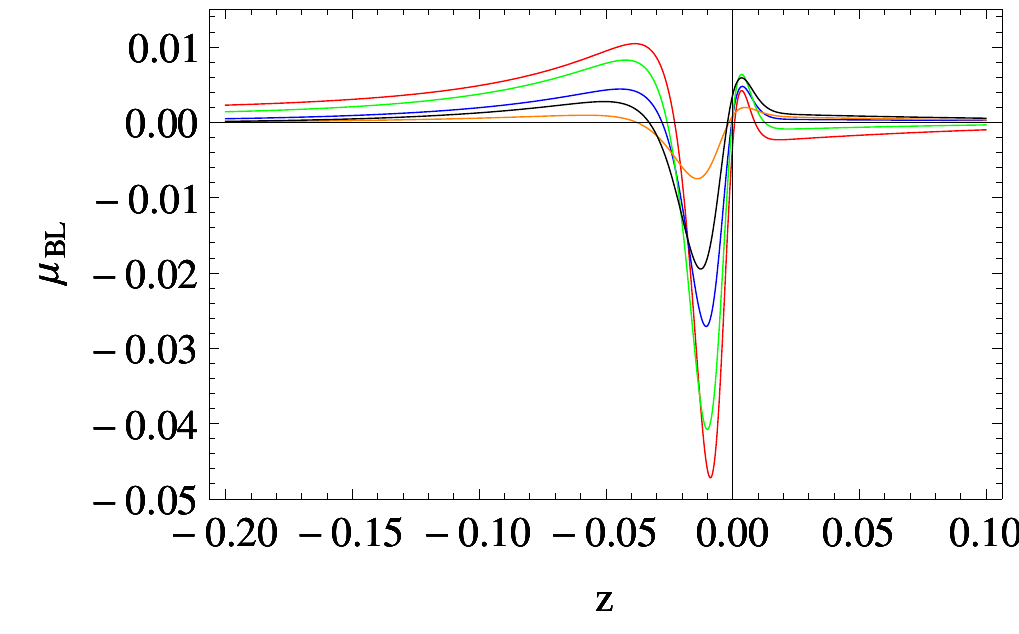}\\
\hspace{0.03\textwidth}
\includegraphics[width=.45\textwidth]{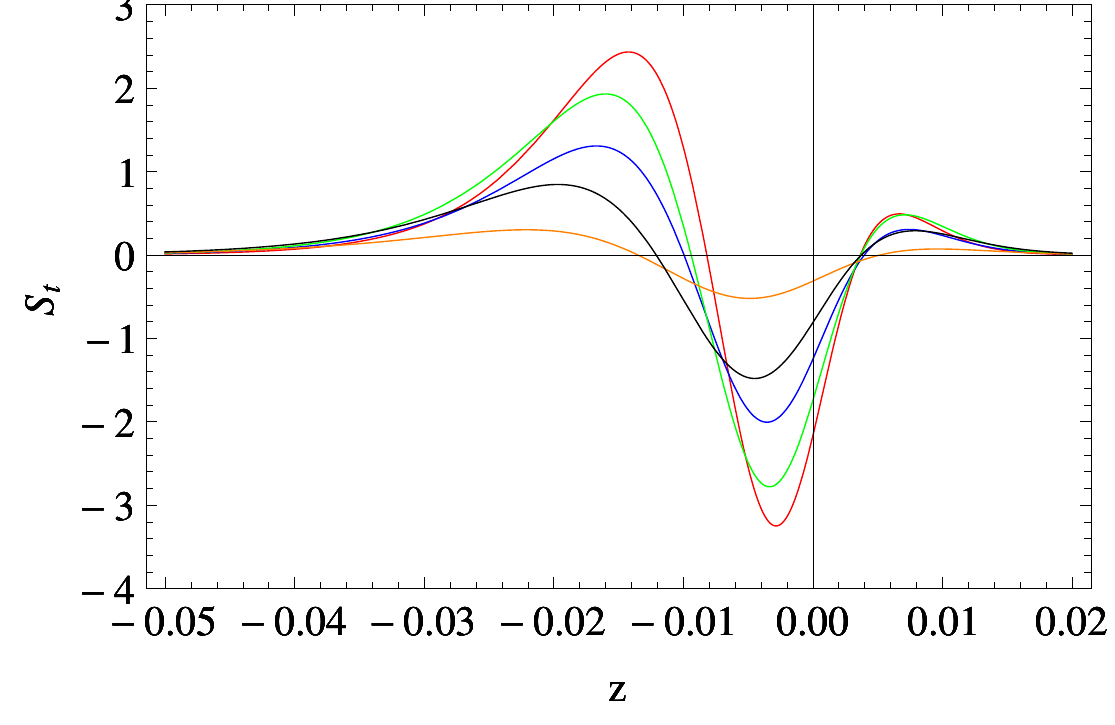}\\
\hspace{0.003\textwidth}
\includegraphics[width=.4\textwidth]{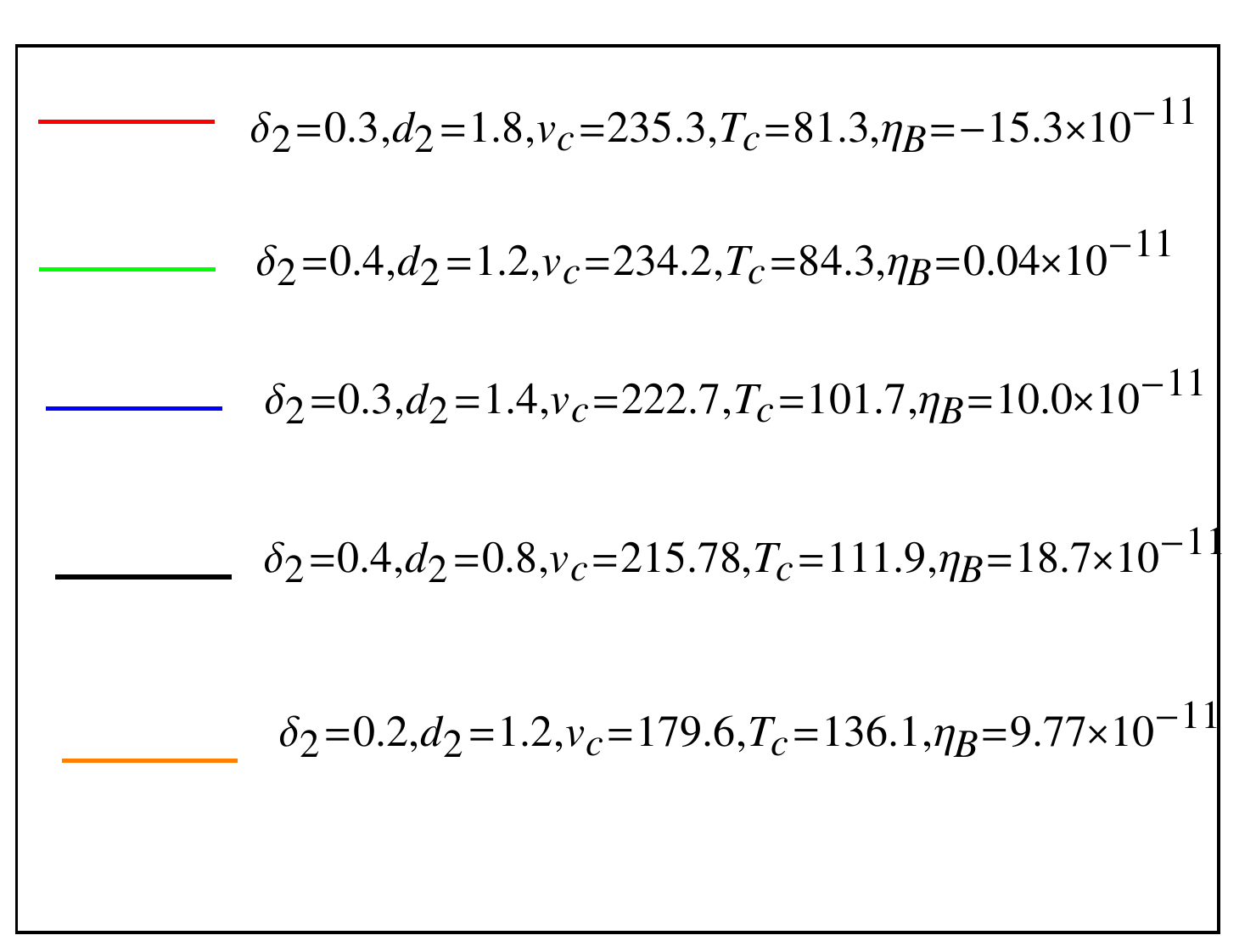}
\caption{ Top and middle panels are the plots of $\mu_{B_L}$ and $S_t$ as a function of the coordinate transversing to the wall ($z$), with benchmark points
being given in the bottom panel.  }
\label{fig:must}
\end{figure}

 As could be observed in the two panels of  Fig.~\ref{fig:baurecalc2}, a larger baryon asymmetry can be generated during the two-step type phase transition process. And, $7\times 10^{-11}<\eta_B<9\times10^{-11}$ turns out to live in the two-step type EWPT favored region for the parameters set.
Last but no least, we would like to go more deeper to the inner mechanism of the sign flip of $\eta_B$ and relevant behaviors with respect to $v_c/T_c$ as shown in Fig.~\ref{fig:baurecalc2}. The BAU magnitude is highly related with that of $\mu_{B_L}$ (see Eq.~\ref{eta1}), which is supposed to be dominated by the behavior of the source term $S_t$ \footnote{Which is the same as that of the $S_\theta$ in~\cite{Fromme:2006wx}, here we denote it as $S_t$ to depict
that it is the top source term. } .
To illustrate the physical picture more clear, we take some
typical benchmark points being used in Fig.~\ref{fig:baurecalc2}, and plot the magnitude of $\mu_{B_L}$ and the dominated source term $S_t$ with
respect to the spacial coordinate $z$, see Fig.~\ref{fig:must}.
The oscillation amplitude of $\mu_{B_L}$ and $S_t$ are found to be increasing
with the increase of $v_c/T_c$. With the increase of the oscillation amplitude of $\mu_{B_L}$ and $S_t$, $\eta_B$ increases
 at first and decreases latter. And the volatility of the oscillation of $\mu_{B_L}$ and $S_t$ give rise to unphysical BAU magnitude, i.e., the negative value of $\eta_B$ appears (see the red line in Fig.~\ref{fig:must}).

\subsection{EDM}

\begin{figure}[!ht]
\centering
\includegraphics[width=.4\textwidth]{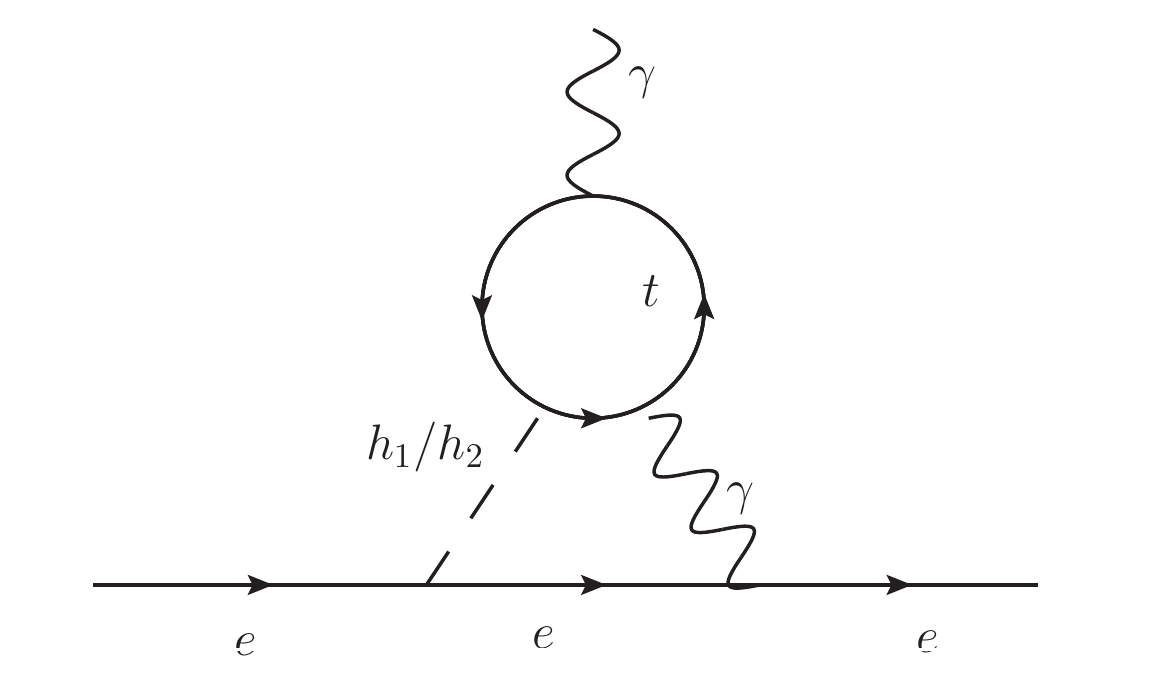}
\caption{Barr-Zee Feynman diagram.}\label{fig:BZ}
\end{figure}

In this section, we use the eEDM search to constrain the CP violation phase. The eEDM
contribution is dominated by the Barr-Zee diagram~\cite{Barr:1990vd}, as illustrated in Fig.~\ref{fig:BZ}.
Where, the two scalar mass eigenstates($h_{1,2}$)
give two opposite sign contributions, obtained
using~\footnote{Here, only the dominate top quark contribution and $h\gamma\gamma$ interaction have been considered. As for the
most general case, and possible cancellation mechanism we refer to~\cite{Bian:2014zka}.  }
\bea
\label{EEDM}
\frac{d_e}{e} &=&  \frac{16}{3}\frac{\alpha}{(4\pi)^3}\frac{m_e}{v^2} \nonumber \\
 && \times \,\frac{bv_s v}{2\Lambda^2}\,Z\,\Big[G\left(m_t^2/m_1^2\right)-G\left(m_t^2/m_2^2\right)\Big]\; ,
\eea
where $m_{e,t}$ are the electron (top quark) masses, and $Z$ characterizes the mixing between $h$ and $S$~\cite{Weinberg:1990me},
being expressed as:
\begin{equation}\label{Z}
Z=\cos\phi\sin\phi=\frac{m_{sh}^2}{\sqrt{(m_h^2-m_S^2)^2+4m_{Sh}^4}} \; ,
\end{equation}
and the function $G(z)$,
\begin{equation}
G(z)\equiv z \int_0^1dx \frac{1}{x(1-x)-z}\log\left[\frac{x(1-x)}{z}\right]\; .
\end{equation}

The magnitude of eEDM, in the parameter regions of Fig.~\ref{fig:baurecalc2}, decreases with the increase of $\delta_1(\delta_2)$ and is below the limit given by the experimental results of ACME~\cite{Baron:2013eja}, which means that the CP violation phase in the parameter spaces is allowed by the eEDM experiment ACME.

\section{Dark matter and Electroweak baryogenesis}

In this section, we investigate the possibility to explain the dark matter relic abundance and EWBG in the same parameter spaces in the model.
To achieve this purpose, we go through the dark matter issues at first. Then we explore the relation between the dark matter mass and the EWPT. The explanation of the the dark matter relic abundance and the EWBG will be acquired simultaneously at last.

\subsection{Dark matter}
\label{relic_dm}

\begin{figure}[!htb]
\centering
\includegraphics[width=.5\textwidth]{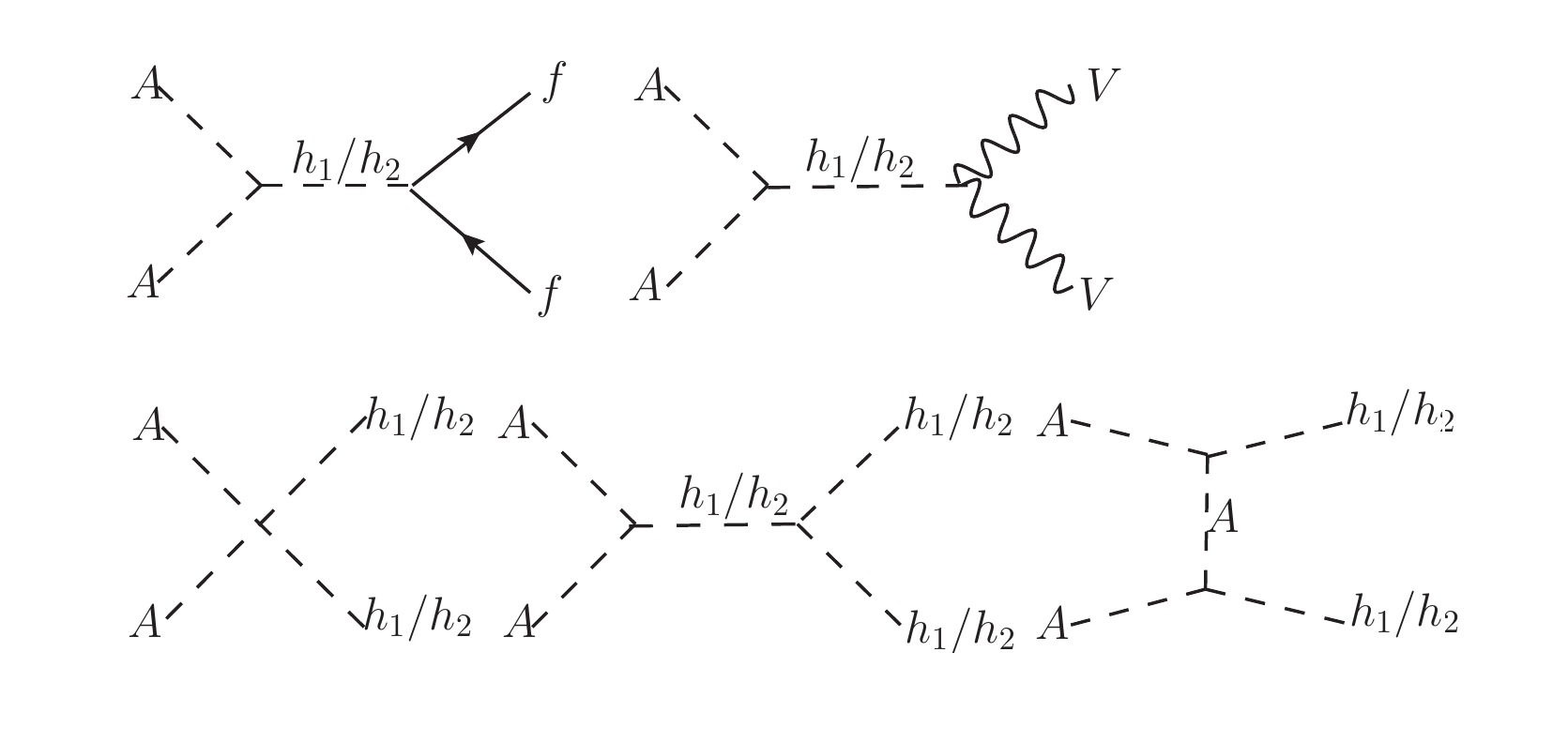}
\caption{Annihilation channels of dark matter particle pairs.}
\label{fig:anhichann}
\end{figure}

In our model, the thermal relic density $\Omega_Ah^2$ is mainly controlled by the pair annihilation cross section of $A$, i.e., $\Omega_A h^2\sim 1/\langle\sigma v\rangle$,
in which $\langle\sigma v\rangle$ is the thermal average of the product of annihilation cross section and relative velocity,
with relevant Feynman diagrams being shown in Fig.~\ref{fig:anhichann}.


Direct detection experiments are searching for dark matter scattering off atomic nuclei, and in the model we only have the Higgs mediated
 spin-independent scattering cross section which constrains our parameter spaces.
The scattering cross section of $A$ with a proton for our model
casts the form of,
\begin{eqnarray}
\sigma_{dd}& =& \frac{m_p^4}{2\pi v^2\of{m_p+m_{A}}^2}
\times \of{\frac{g_{AAh_1}\cos\phi}{m_{h_1}^2}-\frac{g_{AAh_2}\sin\phi}{m_{h_2}^2}}^2 \nonumber\\
&&\times \of{f_{pu}+f_{pd}+f_{ps}+\frac{2}{27}\of{3f_G}}^2\; ,
\label{eq:dd_cxn}
\end{eqnarray}
which is the same with Eq.~(25) in ~\cite{Gonderinger:2012rd} except that the couplings $g_{AAh_1}$ and $g_{AAh_2}$ are different as shown in the Appendix.~B.

\begin{figure}[!htb]
\centering
\includegraphics[width=.4\textwidth]{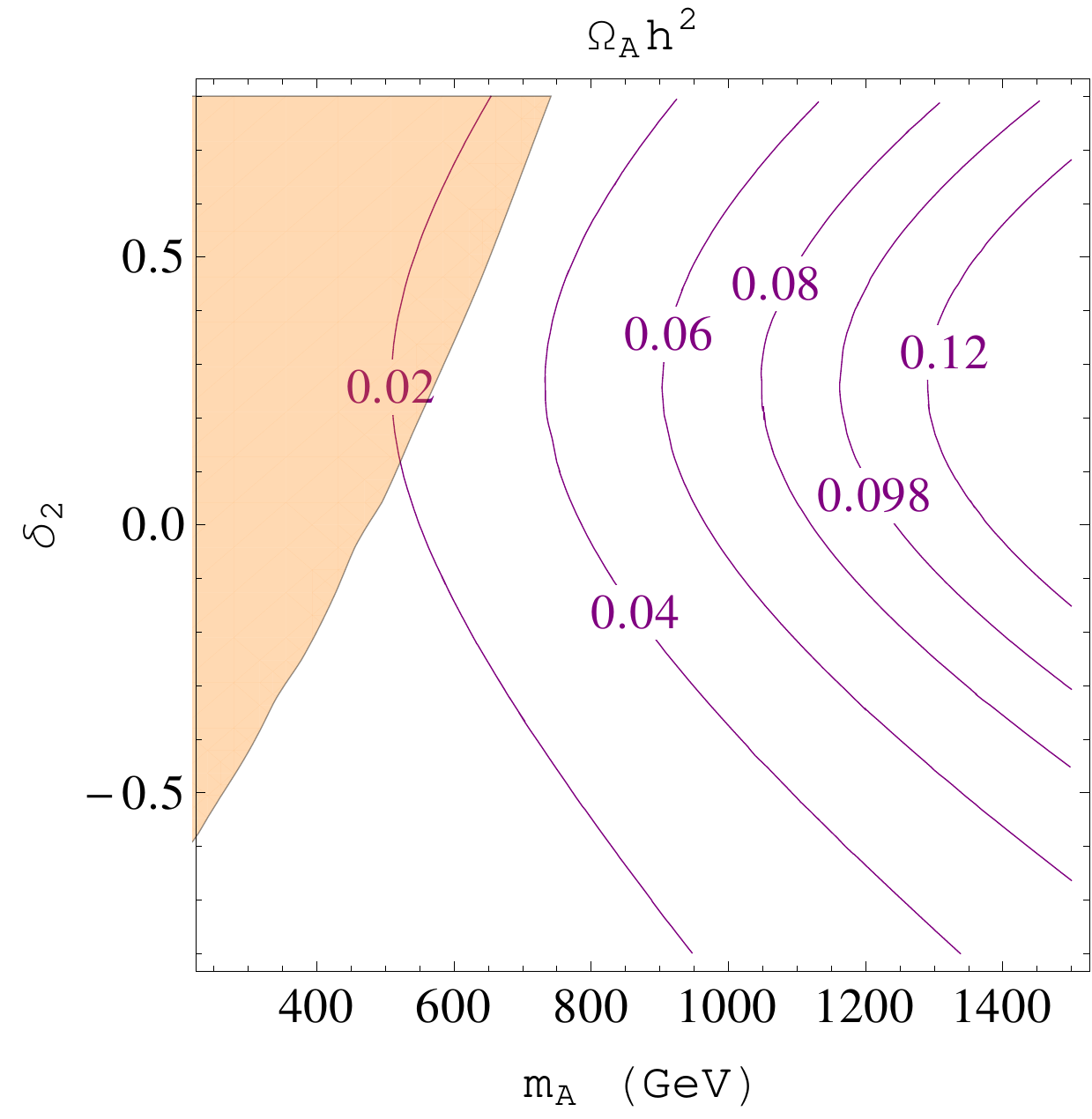}
\includegraphics[width=.4\textwidth]{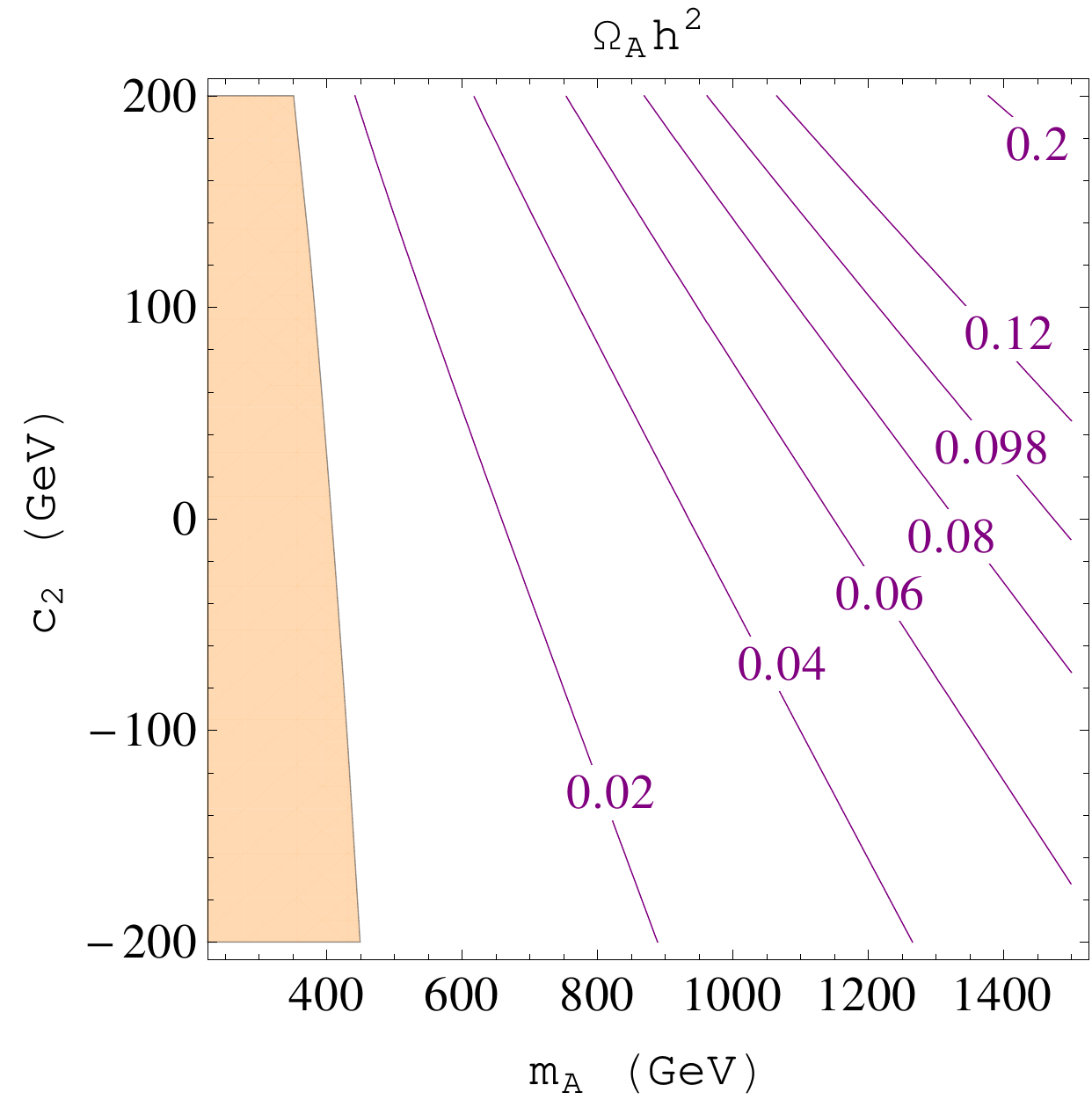}
\caption{Dark matter relic density $\Omega_{A}h^2$ as a function of $c_2$ ($\delta_2$) and $m_A$ with $x=200$ GeV,~$\phi=0.6$, and $m_{h_2}=250$ GeV. The contours of $\Omega_{A}h^2$ are shown by
purple lines with $c_2=30$ GeV ($\delta_2=0.2$) for the top (bottom) panel. The parameter spaces excluded by LUX~\cite{Akerib:2013tjd} are shown by the orange region.
}
\label{fig:dm_delta2}
\end{figure}

Fig.~\ref{fig:dm_delta2} depicts the relic density as a function of $m_A$ and $\delta_2$ ($c_2$). The magnitude of the relic density increases
with the increase of $m_A$, since the annihilation cross section is inversely proportional to the dark matter mass. The LUX experimental results mostly constrain the smaller dark matter masses region, because the $\sigma_{dd}$ has the dark matter mass suppress effect, as could be seen in Eq.~\ref{eq:dd_cxn}.
The $\delta_2$ and $c_2$ affect the magnitude of the relic density and direct detection of the dark matter mainly through the couplings $g_{AAh_1}$
and $g_{AAh_2}$.

\subsection{The dark matter mass and the EWPT}
\label{dmewpt}
In this section, we explore the relation between the dark matter mass and the EWPT.


\begin{itemize}[leftmargin=.5cm,rightmargin=.5cm]

\item The dark matter mass, $T=0$ and $T\neq 0$ vacuum structure

The full analysis of the vacuum structure in our model is not easy, since there are many extreme points in the three dimensional space and the lack of $Z_2$ symmetry of $S$ makes these extreme solutions very complicated.
In the numerical scan of the parameter space in the EWPT part, we find that in most cases there are no local minimum at $A\ne0$, whether at $T=0$ or $T\ne0$, especially when $m_A$ is large.
 This can be understood partly through the following arguments. Suppose there is a local minimum ($v_h$,$v_s$,$v_A$), it should also be a minimum in the {$S=v_s$,$h=v_h$} subspace. In this subspace, the potential $V_0(v_h,v_s,A)$ has a local minimum when the coefficient of $A^2$ term is negative:
 \beq
  \frac{b_2+b_1}{2}+\frac{\sqrt{2}c_2}{6}v_s+\frac{d_2}{4}v_s^2+\frac{\delta_2}{4}v_h^2<0\;.
 \eeq
 Using the Eq.~\ref{replace m2,b2}, we rewire the above condition as
 \beq
 m_A^2+\frac{\sqrt{2}c_2}{6}(x-v_s)+\frac{d_2}{4}(v_s^2-x^2)+\frac{\delta_2}{4}(v_h^2-v^2)<0\; .
 \label{eq:con}
 \eeq
 As a nature expectation, $v_s\sim x$ and $v_h\sim v$. So when $m_A$ is large, the condition of Eq.~\ref{eq:con} can be easily violated and a local minimum at $A\ne0$ does not exist in this subspace, neither in the whole space. At $T\ne0$, thermal contributions stabilize $A$ field at the origin so $A\ne0$ minimum does not exist at high temperatures.


\item EWPT and the dark matter mass

With $A=0$, the Eq.~\ref{eq:potential_tree}  recasts the form of
\begin{eqnarray}
V_0 (h,S) &&= \frac{m^2}{4}h^2 + \frac{\lambda}{16}h^4+\frac{\sqrt{2}}{8}\delta_1h^2S + \frac{\delta_2}{8}h^2 S^2\nonumber\\
 && + \frac{1}{4}\of{b_2-b_1}S^2
+ \frac{\sqrt{2}}{12}c_2 S^3 + \frac{d_2}{16}S^4 \; ,
\end{eqnarray}
Thus, the vacua gap $\Delta V=V_h-V_S$ at tree level could be obtained as,
\begin{eqnarray}
\Delta V &=& - \frac{m^2 v^2}{4}+\frac{1}{4}(b_2-b_1) v_s^2 + \frac{c_2 (v_s^3 - x^3 )}{6\sqrt{2}} + \frac{d_2 (v_s^4 - x^4)}{16}  \nonumber\\
&&- \frac{b_2 - b_1}{4} x^2 - \frac{v^2 x \delta_1}{4 \sqrt{2}}
-\frac{v^2 x^2 \delta_2}{8} -\frac{\lambda}{16} v^4\; ,
\end{eqnarray}
with $v_s$ being the solution of $\partial_S V_0(0,S)=0$.
Since the dark matter mass could only affect the combination $b_1+b_2$, and the combination $b_2-b_1$ is a constant with other parameters being fixed (see Eq.~\ref{replace m2,b2}),
one finds that the dark matter mass is irrelevant to the vacuum energy gap, thus one could expect that the strength of EWPT (inversely proportional to $\Delta V$) won't be affected by which.

\end{itemize}

Based on the above two arguments, we found that the EWPT could be independent of the dark matter mass in the relatively larger dark matter mass regions.

\subsection{EWBG and DM}

The higher magnitude of the relic density requires a relatively larger dark matter mass, about $m_A\sim1000$ GeV.
And in this mass region, as explored in the last section and the section~\ref{baucal}, the EWPT and the generation of BAU is independent of $m_A$.
In order to accommodate the dark matter relic abundance and the EWBG simultaneously we assume the dark matter mass $m_A= 1500$ GeV.

\begin{figure}[!htp]
\centering
\includegraphics[width=.4\textwidth]{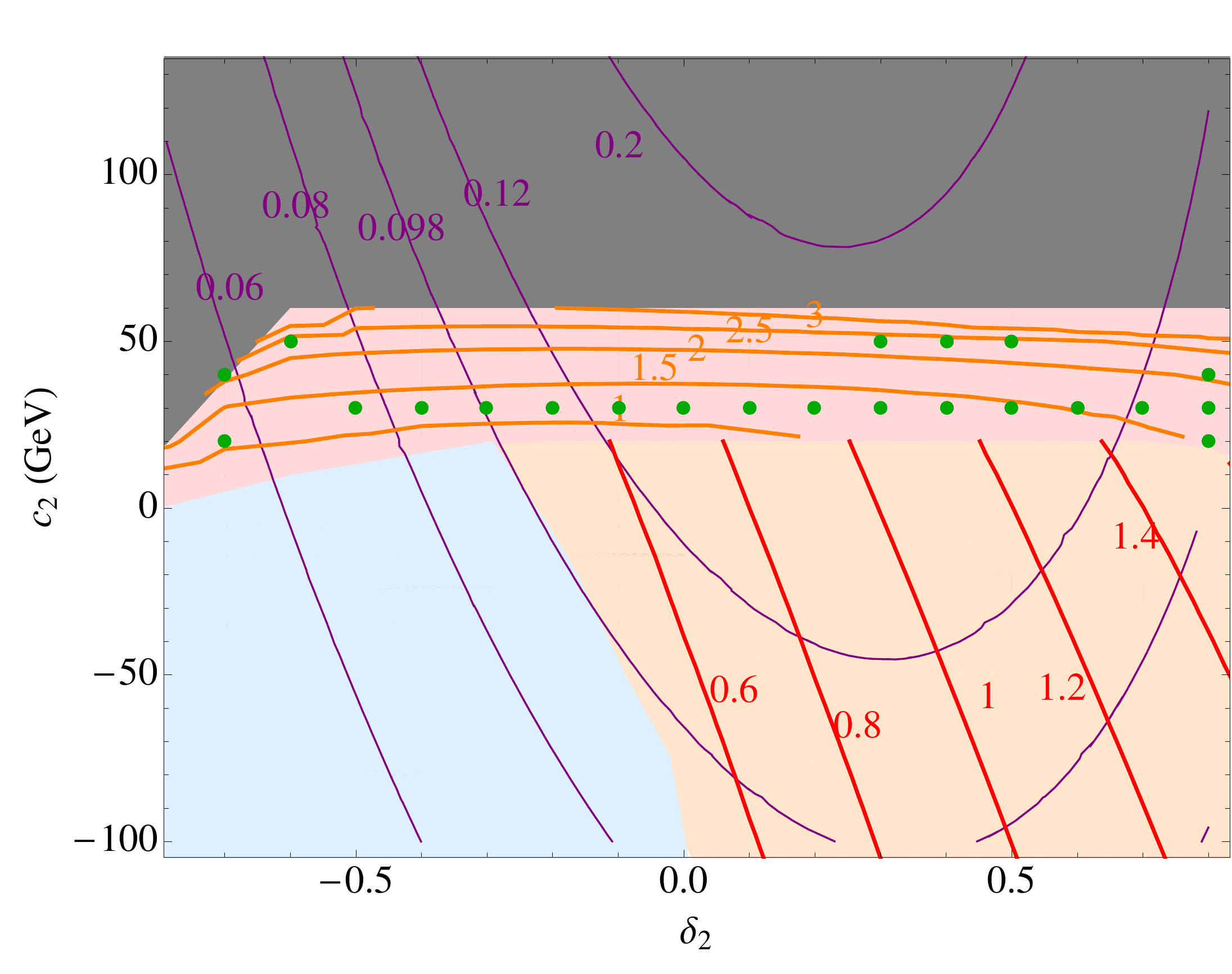}
\caption{The contours of the relic density of the dark matter ($\Omega_{A}h^2$) and the magnitude of $v_c/T_c$ with $x=200$ GeV,~$\phi=0.6$, $m_A=1500$ GeV, and $m_{h_2}=250$ GeV. The contours of $\Omega_{A}h^2$ are shown by
purple lines and the light orange(red) lines depict the value of $v_c/T_c$. The green points are parameter spaces with appropriate BAU magnitude $0.7\times 10^{-10}<\eta_B<0.9\times 10^{-10}$.
}
\label{fig:composite2}
\end{figure}

 We present  in Fig.~\ref{fig:composite2} the combined result of EWBG and DM relic density.
All parameter regions in this plot are under the direct detection constraint from LUX.
Fig.~\ref{fig:composite2} indicates that the behaviors of one- and two-step phase transitions are
dominated by $\delta_2$ and $c_2$ respectively. The dark matter relic density could be affected by $\delta_2$ and $c_2$,
since they enter into the couplings of the dark matter particle and Higgs, i.e., $g_{AAh_1(h_2)}$, which determines the dark matter particle pairs annihilation processes. A larger value of $c_2$ leads to a larger $\Omega_A h^2$, and the correct dark matter relic density observed by~\cite{Jarosik:2010iu,Ade:2013zuv} could be obtained in some parameter spaces of $\delta_2$ and $c_2$.
As for the strength of EWPT and the magnitude of BAU, we found that the two-step SFOEWPT case is more favored.
The magnitude of eEDM is obtained to be $|d_e|= 4\times10^{-29}~ecm$, below
the upper limit given by ACME.

\section{Comments on the calculation methods involving SFOEWPT and BAU}

 As pointed out by \cite{Patel:2011th},
the SFOEWPT criteria $v_c/T_c$ calculated in the traditional mathod is subject to gauge problems, as the $v_c$ and $T_c$ both involve gauge dependence problem. The EWPT is performed in Landau gauge in this paper and one needs to take this into account when interpreting our results. We have checked the robustness of the analysis in the model by following the gauge independent treatment introduced in \cite{Patel:2011th}
for part of the parameter spaces favored by the EWBG and DM
explanations in Fig.~\ref{fig:composite2}. We find that the magnitude of $T_c$ drops a lot and
the corresponding $\bar{v}(T_c)/T_c$ is higher than $v_c/T_c$ for the two-step case.

The numerical calculation of the BAU is based on the source terms and chemical potentials transport equations being computed within the WKB approximation. The results being given might be different to some extent in comparison with the results calculated using the quantum transport equations from the close time path (CTP) formulation, wherein the CP violation source is derived using the ``(vacuum expectation value)vev-insertion" approach~\cite{Lee:2004we,Chung:2009qs,Morrissey:2012db,Kozaczuk:2012xv}.

\section{Discussions and conclusion}

 By supplementing a complex singlet to the SM, DM relic abundance and EWBG could be accommodated simultaneously.
Both two ingredients of EWBG, i.e., the SFOEWPT and new CP violation phase, have been explored at the same time.
The imaginary part of the complex singlet (A) makes the DM candidate due to the reduced $Z_2$ symmetry from
a global $U(1)$ symmetry.

 The phase transition could be one- and two- step types
as found by the study of the energy gap $\Delta V$. In order to reveal the effects of cubic and quartic terms
on the EWPT, we performed EWPT analysis
numerically using {\tt CosmoTransitions} package in the parameter spaces of $(c_2,~\delta_1), (d_2,~\delta_2)$ and $(\delta_2,~c_2)$.
Both one- and two- steps first order phase transition are obtained, and the two-step case favors SFOEWPT more than
that of one-step case. And this study is consistent with the analysis of $\Delta V$, i.e., a lower magnitude of $\Delta V$ results in a
larger strength of EWPT.

The newly introduced CP violation makes the EWBG mechanism viable, and the CP violation phase is found to be allowed by constraints of ACME.
The general and specific calculations of $\eta_B$ shows that the magnitude of BAU is highly related with $v_c$ and $T_c$,
and the increase of $v_c/T_c$ does not always cause the increase of $\eta_B$, because $v_c/T_c$ affects the CP violation source.

The magnitude of the relic density of the DM is found to increase with the increase of the dark matter mass.
The smaller DM mass regions are severely constrained by direct detection
experimental results. And the EWPT is irrelevant to the DM mass in the larger DM mass regions.
At last, we present a scenario in which both correct DM relic density
and successful EWBG could be achieved in some parameter spaces of $\delta_2$ and $c_2$ considering the eEDM experiment ACME.

There is a possibility that baryogenesis and dark matter are generated at the time of EWPT simultaneously with the dark matter sector being asymmetric dark matter~\cite{Servant:2013uwa,Allahverdi:2013tca,Cheung:2013dca},
thus shed light on the path to explain the ratio between the observed abundance of DM and baryons in the present day: $\Omega_{DM}/\Omega_B\approx5.4$
~\cite{Ade:2013zuv}. The key ingredient for the test of this kind of model is the colliders search of the Higgs triple coupling~\cite{Curtin:2014jma,Katz:2014bha,Craig:2014lda,Craig:2013xia,Profumo:2014opa,Henning:2014gca} and the introduced CP violation~\cite{He:2014xla,Dolan:2014upa,Shu:2013uua,Draper:2009au,Gupta:2009wu}, that
will be postponed to the future works.

\appendix
\section{Field dependent masses}

Field dependent masses matrix is given by,
\begin{equation}
m_{h,S,A}^2 = eigenvalues\of{ M^2}\; ,
\end{equation}
with
\begin{eqnarray}
M^2=\lb\begin{array}{ccc} m_{hh}^2\of{h,S,A} & m_{hS}^2\of{h,S,A} & m_{hA}^2\of{h,S,A} \\  m_{hS}^2\of{h,S,A} & m_{SS}^2\of{h,S,A} & m_{SA}^2\of{h,S,A}\\m_{hA}^2\of{h,S,A}&m_{SA}^2\of{h,S,A}& m_{AA}^2\of{h,S,A}\end{array})\rb\; ,\nonumber
\end{eqnarray}
and matrix entries are given by
\begin{eqnarray}
m^2_{hh}&=&\frac{m^2}{2}+\frac{\sqrt{2}\delta_1}{4}S+\frac{\delta_2}{4}\of{S^2+A^2}+\frac{3\lambda}{4}h^2\;,
\\
m^2_{hS}&=&\frac{\sqrt{2}\delta_1}{4}h+\frac{\delta_2}{2}h S\; ,\\
m^2_{SS}&=&\frac{-b_1+b_2}{2}+\frac{d_2}{4}A^2+\frac{\sqrt{2}c_2}{6}S+\frac{3 d_2}{4}S^2\nonumber\\
&&+\frac{\delta_2}{4}h^2\; ,\\
m^2_{hA}&=&\frac{\delta_2}{2}hA\;, \\
m^2_{SA}&=&\frac{\sqrt{2}c_2}{6}A+\frac{d_2}{2}A S\;, \\
m^2_{AA}&=&\frac{b_1+b_2}{2}+\frac{3d_2}{4}A^2+\frac{\sqrt{2}c_2}{6}S+\frac{d_2}{4}S^2\nonumber\\
&&+\frac{\delta_2}{4}h^2\; ,
\end{eqnarray}
and other field dependent masses given by,
\begin{eqnarray}
&&m_{G}^2 = \frac{m^2}{2}+\frac{\sqrt{2}\delta_1}{4}S+\frac{\delta_2}{4}\of{S^2+A^2}+\frac{\lambda}{4}h^2\; ,\\
&&m^2_{t}=\frac{1}{2}y_t^2h^2\; ,\\
&&m^2_{W^\pm}=\frac{1}{2}g^2h^2\; ,\\
&&m^2_{Z}=\frac{1}{2}\of{g^2+g'^2}h^2\; ,
\end{eqnarray}
with $G$ being goldstone fields.

\section{Scalar triple and quartic couplings }
\begin{eqnarray}
g_{AAh_1} &=& -2i(\delta_2v\cos\phi/4 + (\sqrt{2}c_2/12 + d_2x/4)\sin\phi)\; ,\nonumber\\
g_{h_1h_1h_1} &=&-6i(\lambda v\cos^3\phi/4 + (d_2x/4 + \sqrt{2}c_2/12))\sin^3\phi\;  \nonumber\\
&&+ (\delta_2x/4 + \sqrt{2}\delta_1/8)\cos^2\phi\sin\phi\nonumber \\
&& + \delta_2v\cos\phi\sin^2\phi/4)\; ,\nonumber \\
g_{AAh_2} &=&-2i(-\delta_2v\sin\phi/4 + (\sqrt{2}c_2/12 + d_2x/4)\cos\phi)\;  ,\nonumber\\
g_{h_1h_1h_2} &=& -2i(-3\lambda v\cos^2\phi\sin\phi/4 + 3\sin\phi^2\cos\phi(d_2x/4 \nonumber\\
&&+ \sqrt{2}c_2/12) - (\delta_2x/2 + \sqrt{2}\delta_1/4)\cos\phi\sin^2\phi \nonumber\\
&&+ \delta_2v/2\sin\phi\cos^2\phi)\;, \nonumber\\
g_{AAh_1h_1} &=&-4i(\delta_2\cos^2\phi /8+ d_2\sin^2\phi/8)\;,\nonumber \\
g_{h_2h_2h_2} &=& -6i(-\lambda v\sin^3\phi/4 + (d_2x/4 + \sqrt{2}c_2/12))\cos^3\phi \nonumber\\
&&+ (\delta_2x/4 + \sqrt{2}\delta_1/8)\cos\phi\sin^2\phi\nonumber\\
&& - \delta_2v\sin\phi\cos^2\phi/4)\;,\nonumber \\
g_{AAh_2h_2} &=& -4i(\delta_2\sin^2\phi/8 + d_2\cos^2\phi/8)\; ,\nonumber\\
g_{h_2h_2h_1} &=&-2i(3\cos^2\phi\sin\phi(d_2x/4 + \sqrt{2}c_2/12)\nonumber\\
&& + 3\sin^2\phi\cos\phi\lambda v/4 - (\delta_2x/2 + \sqrt{2}\delta_1/4) \nonumber\\
&&\times\cos^2\phi\sin\phi
- \delta_2v\sin^2\phi\cos\phi/2)\;, \nonumber\\
g_{AAh_1h_2} &=&-2i(\sin\phi\cos\phi(d_2-\delta_2)/4)\; .
\end{eqnarray}
\section{Thermal masses used in EWPT analysis}

Thermal masses of bosonic field relevant to EWPT study are listed bellow,
\begin{eqnarray}
&&\Pi_{G}=(3 g^2/16+g'^2/16+\lambda /8+y_t^2/4+\delta_2 /24)T^2\;,\nonumber \\
&&\Pi_{hh}=(3g^2/16+g'^2/16+\lambda /8+y_t^2/4+\delta_2/24)T^2\;,\nonumber \\
&&\Pi_{SS}=(d_2/12+\delta_2 /48)T^2\;,\nonumber \\
&&\Pi_{AA}=(d_2/12+\delta_2 /48)T^2\; ,\nonumber \\
&&\Pi_{W}^L=\Pi_{Z}^L=11g^2/6 T^2\;\;\nonumber \\
&&\Pi_{W}^T=\Pi_{Z}^T=0\; .
\end{eqnarray}

\section{Transport equations  }
For the case when the top quark acquires a complex mass during the electroweak phase transition, $m=|m(z)|e^{i\theta(z)}$, the source terms and transport equations were derived in Ref.~\cite{Fromme:2006wx} using the WKB approximation . Here we briefly summarize the approach with notations adopted in Ref.~\cite{Fromme:2006wx}. With $v_w$ denoting the wall velocity, the plasma velocity $u$ and the coefficients $K_i$, $\tilde K_i$ are defined by thermal averages. And the chemical potentials are expanded to second order in gradients.

The source terms induced by the top quark are
\bea \label{source}
S_{\mu}&=&K_7\theta'm^2 \mu_1'\; ,
\nonumber\\
S_{\theta}&=&-v_wK_8(m^2\theta')'+v_wK_9 \theta'm^2(m^2)'\; ,
\nonumber\\
S_{u}&=&-\tilde K_{10}m^2\theta'u_1'\; .
\eea
in which the primes denotes derivatives with respect to $z$ (the direction perpendicular to the wall) and the CP-even first order perturbations $u_1$, $\mu_1$ can be solved by
\bea
\label{1}
&&v_w K_1 \mu_1' + v_w K_2 (m^2)' \mu_1 + u_1'  -\Gamma^{\rm tot}\mu_{1}
=v_wK_{3}(m^2)'\; ,
\nonumber\\[.3cm]
&&-K_4 \mu_1' + v_w\tilde K_5 u_1' + v_w\tilde K_6 (m^2)'u_1+\Gamma^{\rm tot}u_1
=0\; .\nonumber
\eea
And, the $S_{\theta}$ is the dominated one, effects induced by $S_{u}$ and $S_{\mu}$ are
negligible.

With the
weak sphaleron rate \cite{Mws}, the strong sphaleron rate \cite{Mss}, the top
Yukawa rate \cite{HN95}, the top helicity flip rate, the Higgs number violating
rate \cite{HN95}, the W scattering rate, the quark diffusion constant \cite{JPT95} and
the Higgs diffusion constant  \cite{CJK00}  being given by,
\begin{eqnarray} \label{rates}
&&\Gamma_{ws}=1.0\times10^{-6}T, \quad  \quad \Gamma_{ss}=4.9\times10^{-4}T,
\nonumber \\
&&\Gamma_y=4.2\times10^{-3}T, \quad \quad ~~\Gamma_m=\frac{m_t^2(z,T)}{63T},
\nonumber \\
&&\Gamma_h=\frac{m_W^2(z,T)}{50T},\quad \qquad~~\Gamma_W=\Gamma^{\rm tot}_h,\nonumber\\
&&D_q=\frac{6}{T},\qquad\qquad\qquad~~ D_h=\frac{20}{T}\; ,
\end{eqnarray}
the transport equations of chemical potentials, i.e., left-handed SU(2) doublet tops $\mu_{t,2}$,  left-handed
SU(2) doublet bottoms $\mu_{b,2}$, left-handed SU(2) singlet tops $\mu_{t^c,2}$(charge conjugation of $t_R$),
Higgs bosons $\mu_{h,2}$, are given by,
\bea
&&
-3K_{4,b}\mu_{b,2}'+3v_w\tilde K_{5,b}u_{b,2}'+3\Gamma^{\rm tot}_bu_{b,2}=0\; ,
\nonumber\\[0.5cm]
&&-2K_{4,h}\mu_{h,2}'+2v_w\tilde K_{5,h}u_{h,2}'+2\Gamma^{\rm tot}_hu_{h,2}=0\;,\nonumber\\[0.5cm]
&&2v_wK_{1,h}\mu_{h,2}'+2u_{h,2}'
-3\Gamma_y(\mu_{t,2}+\mu_{b,2}+2\mu_{t^c,2}+2\mu_{h,2})\nonumber\\[0.5cm]
&&-2\Gamma_h\mu_{h,2}=0\; ,
\eea
\begin{widetext}
\bea \label{mus}
&&3v_wK_{1,t}\mu_{t,2}'+3v_wK_{2,t}(m_t^2)'\mu_{t,2}+3u_{t,2}'-3\Gamma_y(\mu_{t,2}+\mu_{t^c,2}+\mu_{h,2})-6\Gamma_m(\mu_{t,2}+\mu_{t^c,2})
-3\Gamma_W(\mu_{t,2}-\mu_{b,2})
\nonumber\\
&&-3\Gamma_{ss}[(1+9K_{1,t})\mu_{t,2}+(1+9K_{1,b})\mu_{b,2}+(1-9K_{1,t})\mu_{t^c,2}]=3K_{7,t}\theta_t'm_t^2\mu_{t,1}'\; ,
\nonumber\\[0.5cm]
&&3v_wK_{1,b}\mu_{b,2}'+3u_{b,2}'-3\Gamma_y(\mu_{b,2}+\mu_{t^c,2}+\mu_{h,2})
-3\Gamma_W(\mu_{b,2}-\mu_{t,2})
-3\Gamma_{ss}[(1+9K_{1,t})\mu_{t,2}\nonumber\\
&&+(1+9K_{1,b})\mu_{b,2}+(1-9K_{1,t})\mu_{t^c,2}]
=0\; ,
\nonumber\\[0.5cm]
&&3v_wK_{1,t}\mu_{t^c,2}'+3v_wK_{2,t}(m_t^2)'\mu_{t^c,2}+3u_{t^c,2}'-3\Gamma_y(\mu_{t,2}+\mu_{b,2}+2\mu_{t^c,2}+2\mu_{h,2})-6\Gamma_m(\mu_{t,2}+\mu_{t^c,2})\; ,
\nonumber\\
&&-3\Gamma_{ss}[(1+9K_{1,t})\mu_{t,2}+(1+9K_{1,b})\mu_{b,2}+(1-9K_{1,t})\mu_{t^c,2}]=3K_{7,t}\theta_t'm_t^2\mu_{t^c,1}'\; ,
\nonumber\\[0.5cm]
&&-3K_{4,t}\mu_{t,2}'+3v_w\tilde K_{5,t}u_{t,2}'+3v_w\tilde K_{6,t}(m_t^2)'u_{t,2}
+3\Gamma^{\rm tot}_tu_{t,2}
=-3v_wK_{8,t}(m^2_t\theta_t')'+3v_wK_{9,t} \theta_t'm^2_t(m_t^2)'\nonumber\\
&&\qquad\qquad\qquad\qquad\qquad\qquad\qquad\qquad\qquad\qquad\qquad\qquad\qquad\qquad
-3\tilde K_{10,t}m^2_t\theta_t'u_{1,t}'\; ,\nonumber\\[0.5cm]
&&
-3K_{4,t}\mu_{t^c,2}'+3v_w\tilde K_{5,t}u_{t^c,2}'
+3v_w\tilde K_{6,t}(m_t^2)'u_{t^c,2}+3\Gamma^{\rm tot}_tu_{t^c,2}=-3v_w K_{8,t}(m^2_t\theta_t')'+3 v_w K_{9,t} \theta_t'm^2_t(m_t^2)'\nonumber\\
&&\qquad\qquad\qquad\qquad\qquad\qquad\qquad\qquad\qquad\qquad\qquad\qquad\qquad\qquad-3\tilde K_{10,t}m^2_t\theta_t'u_{1,t^c}'\; .
\eea
\end{widetext}

\acknowledgments
LGB thank Michael Ramsey-Musuolf for useful discussions.

\end{document}